\documentclass[a4paper,fleqn,usenatbib]{mnras}

\usepackage{newtxtext,newtxmath}

\usepackage[T1]{fontenc}
\usepackage{ae,aecompl}

\usepackage{graphicx}	
\usepackage{amsmath}	
\usepackage{amssymb}	
\usepackage{subfigure}
\usepackage{lineno}

\title[Intrinsic alignments and shear estimates]{Measuring the scale dependence of intrinsic alignments using multiple shear estimates}

\author[C. D. Leonard et al.]{
C. Danielle Leonard,$^{1}$\thanks{E-mail: danielll@andrew.cmu.edu}
Rachel Mandelbaum$^{1}$
\newauthor{(The LSST Dark Energy Science Collaboration)} \\ 
$^{1}$McWilliams Center for Cosmology, Department of Physics, Carnegie Mellon University, Pittsburgh, PA 15213, USA
}

\date{Accepted 2018 May 30. Received 2018 May 18; in original form 2018 February 22}

\pubyear{2018}

\begin{document}
\label{firstpage}
\pagerange{\pageref{firstpage}--\pageref{lastpage}}
\maketitle

\begin{abstract}
We present a new method for measuring the scale dependence of the intrinsic alignment (IA) contamination to the galaxy-galaxy lensing signal, which takes advantage of multiple shear estimation methods applied to the same source galaxy sample. By exploiting the resulting correlation of both shape noise and cosmic variance, our method can provide an increase in the signal-to-noise of the measured IA signal as compared to methods which rely on the difference of the lensing signal from multiple photometric redshift bins. For a galaxy-galaxy lensing measurement which uses LSST sources and DESI lenses, the signal-to-noise on the IA signal from our method is predicted to improve by a factor of $\sim 2$ relative to the method of \cite{Blazek2012}, for pairs of shear estimates which yield substantially different measured IA amplitudes and highly correlated shape noise terms. We show that statistical error necessarily dominates the measurement of intrinsic alignments using our method. We also consider a physically motivated extension of the \cite{Blazek2012} method which assumes that all nearby galaxy pairs, rather than only excess pairs, are subject to IA.  In this case, the signal-to-noise of the method of \cite{Blazek2012} is improved.
\end{abstract}




\section{Introduction}
\label{sec:introduction}
\noindent
As light from distant galaxies propagates through space, its trajectory is modified by the presence of massive structures, and observed galaxy images are distorted in an effect known as gravitational lensing.  Weak gravitational lensing -- the case in which the distortion is small and detectable only via averaging over many galaxy images -- is a key cosmological observable of several upcoming surveys, including the Large Synoptic Survey Telescope \citep[LSST;][]{LSSTScienceBook2}, Euclid \citep{EuclidRedBook}, and the Wide-Field Infrared Survey Telescope \citep[WFIRST;][]{WFIRSTReport}. The weak lensing measurements of these surveys are expected to enhance our understanding of the evolution of dark energy, the nature of gravity on cosmological scales, and other fundamental cosmological questions (see, for example, \citealt{Weinberg2013}). Because of the considerable decrease in statistical uncertainties expected for these next-generation weak lensing measurements, it is crucial that we understand and mitigate all systematic effects that may contaminate weak lensing observables (for a review of these effects, see \citealt{Mandelbaum2017}).

Weak gravitational lensing studies typically measure two-point correlations between the shapes of source galaxies (cosmic shear) and/or between the shapes of source galaxies and the positions of foreground lens galaxies, which we call galaxy-galaxy lensing (see, for example, \citealt{vanUitert2017, DES2017}). It is the latter of these, galaxy-galaxy lensing, which we consider in this work. Translating these measurements into cosmological constraints relies on accounting for the sub-dominant levels of correlation which arise due to other effects. In this paper, we focus on the correlation in alignment due to local gravitational effects. These astrophysical correlations are referred to as intrinsic alignments (IA). For a thorough introduction to this phenomenon, see \cite{Troxel2015, Joachimi2015, Kirk2015, Kiessling2015}.

A common approach to dealing with this effect is to marginalize over the parameters of an IA model (for two recent examples of this approach, see \citealt{vanUitert2017, DES2017}). Popular choices include the linear alignment model \citep{Catelan2001,Hirata2004}, which assumes that alignment is `frozen in' at early time and therefore that the IA two-point function is proportional to the linear matter power spectrum, and the related nonlinear alignment model \citep{Bridle2007} which replaces the linear matter power spectrum with its nonlinear counterpart in an attempt to account for late-time growth of structure. However, these models, while reasonably good descriptions on larger scales, are unable to describe the intrinsic alignment correlation in the one-halo regime. Modelling and measuring IA on these smaller scales is an active field of research interest (see, for example, \citealt{Schneider2010, Blazek2012, Singh2014, Sifon2015, Chisari2015, Blazek2015, Blazek2017}), which has yet to result in a universally accepted and fully coherent model. Therefore, any insight into the scale dependence of IA is of great value, as it enables the construction of improved models and thus the more effective mitigation of this systematic effect in weak lensing measurements.

Furthermore, existing methods for empirically constraining the IA contribution to a given galaxy-galaxy lensing measurement generally require a robust way to understand the photometric redshift (photo-z) error distribution of source galaxies (\citealt{Blazek2012, Chisari2014}), such as by obtaining spectroscopic redshifts of a representative subsample of source galaxies or employing cross-correlation techniques. This may be challenging for upcoming lensing surveys, which will image fainter source galaxies than ever before. It is important to quantify, for a given method of measuring intrinsic alignments, the degree to which this source of systematic uncertainty will impact IA mitigation in upcoming surveys.

Recently, it was shown explicitly in \cite{Singh2016} that the use of different galaxy shape-measurement methods results in a scale-independent multiplicative change in the measured amplitude of the intrinsic alignment contribution to the galaxy-galaxy lensing signal. For the three shape-measurement methods examined there, the difference in the measured IA amplitude was on the order of tens of percent. After a series of tests which ruled out point spread function-related systematic errors as well as environmentally-dependent galaxy ellipticity gradients, the suggested explanation for this result was the presence of isophotal twisting, in which the outer radial parts of a galaxy are more aligned with the tidal field than are the inner regions. Different shear estimates are sensitive to different radial separations from the center of source images. Thus, if isophotal twisting is present, it is expected that shear estimates with sensitivity to outer regions will result in a larger intrinsic alignment signal than those with sensitivity to inner regions. This effect was earlier discussed in \cite{Schneider2013}, within the context of a study of radial alignments in the Galaxy and Mass Assembly Survey. It had also been seen in simulation studies of galaxy ellipticities and intrinsic alignments \citep{Tenneti2014, Tenneti2015, Velliscig2015, Velliscig2015b, Hilbert2017}. Following its direct observational detection in \cite{Singh2016}, \cite{Chisari2016} proposed exploiting the effect to probe primordial non-Gaussianity, which is theorised to introduce deviations to the IA signal on large scales, where IA modelling is best-understood.

In this work, we take advantage of the finding of \cite{Singh2016} to construct a new method for measuring the scale dependence of IA contamination to galaxy-galaxy lensing on scales at which nonlinear and one-halo effects dominate.  We consider the difference between two tangential-shear measurements using the same set of source and lens galaxies, differing only in the shear estimation method applied to sources. The lensing contribution to these signals is identical, and is thus removed by taking their difference. However, if the shear estimation methods selected are sensitive to different radial regions of the galaxy light profile, an IA portion of the signal will remain, resulting in a method to determine the IA contribution up to a constant factor and hence to measure its scale dependence. 

For this cancellation of the lensing signal to occur, the source sample associated with both shear estimates must be the same, and any residual multiplicative bias must be sub-dominant. For this reason, we suggest the use of two shear estimates of the Bayesian Fourier Domain (BFD) type (\citealt{Bernstein2014}, \citealt{Bernstein2016}), adjusted to accommodate different radial weighting functions. BFD has been shown to result in sub-dominant multiplicative bias \citep{Bernstein2014}, and the use of two such similar estimates would prevent issues of different source selection cuts.

Because in our method the lensing contribution to the signal is entirely subtracted off, it does not need to be measured and then removed, meaning that this method may be especially robust to challenges in constraining the source galaxy photometric redshift errors. This method also has the potential to reduce the statistical uncertainty in the measured intrinsic alignment signal, due to the respective correlation both between the shape noise of measurements made with different shear estimates, and between the cosmic variance of those measurements. It therefore has the potential to test our small-scale alignments models in a way that results in improved data-driven models. We will investigate and quantify both of these possible advantages of this method.

This paper is organised as follows. In Section \ref{sec:theory} we provide a brief theoretical review of the relevant galaxy-galaxy lensing observables and how they are expected to be affected by IA, then we introduce an existing method for measuring the IA contribution to galaxy-galaxy lensing observables to which we will compare (\citealt{Blazek2012}, hereafter B2012). We proceed, in Section \ref{sec:newmethod}, to present our new method for measuring IA, and we provide details on its implementation. In Section \ref{sec:obsscen}, we describe the two observational scenarios in which we will forecast the capabilities of our proposed method in comparison with the above-mentioned existing method. We present our main results in Section \ref{sec:results}, in which we first describe a modification to the method of B2012 which permits fair comparison with our method. We then consider whether our method improves upon the existing method in terms of robustness to systematic uncertainties, and finally we demonstrate the power of our method in the regime in which statistical errors dominate. We discuss our findings and conclude in Section \ref{sec:conclusions}. Throughout this work, unless otherwise noted, we assume cosmological parameters defined by the Planck 2015 results (\citealt{Planck2015}): $h=0.67$, $\Omega_{\rm c}=0.27$, $\Omega_{\rm b}=0.049$, $A_s=2.2\times10^{-9}$ ($\sigma_8 = 0.84$), with $\Omega_{\rm k}=0$.

\section{Theoretical background}
\label{sec:theory}
\noindent
In this section, we briefly review the theoretical basis for relevant galaxy-galaxy lensing quantities, and discuss the expected intrinsic alignment contribution to the galaxy-galaxy lensing signal. We then describe existing methods for measuring this contribution, focussing on the method of B2012 which will be used in this work as a benchmark against which to measure the performance of our new method.

\subsection{Galaxy-galaxy lensing}
\label{subsec:ggl_theory}
\noindent
Galaxy-galaxy lensing studies are concerned with the measurement of the cross-correlation between the shapes of background source galaxies and positions of foreground lens galaxies. Typically the measured quantity is either $\tilde{\gamma}_t(r_p)$, the average tangential shear of source galaxies about lens galaxies, or $\widetilde{\Delta\Sigma}(r_p)$, the differential projected surface mass density around lens galaxies (where $r_p$ is the projected radial distance from a lens galaxy center in comoving coordinates and a tilde indicates an observed quantity). For a single lens-source pair, $\gamma_t$ and $\Delta \Sigma$ are related via:
\begin{linenomath*}
\begin{equation}
\gamma_t = \frac{\Delta \Sigma}{\Sigma_c },
\label{gam_DS}
\end{equation}
\end{linenomath*}
where $\Sigma_c$ is the critical surface density, which depends on the separation of lens and source galaxies. It is given (in comoving coordinates) by:
\begin{linenomath*}
\begin{equation}
\Sigma_c = \frac{c^2}{4\pi G}\frac{\chi_s}{\chi_l (\chi_{s} - \chi_l)(1 + z_l)},
\label{SigmaC}
\end{equation}
\end{linenomath*}
where $\chi_s$ and $\chi_l$ are the comoving radial distances from observer to source and from observer to lens respectively. $z_l$ is the lens redshift. 
$\Delta \Sigma(r_p)$ is defined as:
\begin{linenomath*}
\begin{align}
\Delta \Sigma(r_p) &=  \overline{\Sigma}(r_p)  - \Sigma(r_p) \nonumber \\ &=
\frac{2}{r_p^2} \int_0^{r_p} dr_p^\prime r_p^\prime \Sigma(r_p^\prime) - \Sigma(r_p), 
\label{DelSigDef}
\end{align}
\end{linenomath*}
where $\Sigma(r_p)$ is the projected surface density of matter about a lens galaxy and $\overline{\Sigma}(r_p)$ is the same quantity averaged within projected radial separation $r_p$. Where required, we will compute the theoretical value of $\Sigma(r_p)$ via
\begin{linenomath*}
\begin{equation}
\Sigma(r_p) = \rho_{\rm m} \int d\Pi^\prime \left(1+\xi_{lm}\left(\sqrt{r_p^2 + (\Pi^\prime)^2}\right)\right),
\label{Sigma}
\end{equation}
\end{linenomath*}
where $\rho_{\rm m}$ is the matter density in comoving units, $\Pi$ is the line-of-sight separation, and we use $\xi_{lm}$ to denote the correlation function of matter with lens galaxies. $\xi_{lm}$ comprises a one-halo and two-halo term, the ingredients for each of which we now discuss.

We first compute the two-halo term of the matter power spectrum using {\tt CLASS} \citep{Lesgourges2011} with {\tt halofit} (\citealt{Smith2003}, \citealt{Takahashi2012}) via {\tt CCL} \citep{CCLrepo} to obtain the nonlinear matter power spectrum, which we then Fourier transform using {\tt FFTlog} \citep{Hamilton2000} to obtain $\xi_{mm}(r)$. We assume a linear galaxy bias $b$ for the lens galaxy sample to convert to $\xi_{lm}(r)$ via $\xi_{lm}(r) = b r_\text{cc}\, \xi_{mm}(r)$, where we set the cross-correlation coefficient $r_\text{cc}$ to unity. In computing the one-halo contribution, we assume a Navarro-Frenk-White (NFW) profile \citep{Navarro1997}, where we follow \cite{Mandelbaum2008} and take the concentration-mass relation to be
\begin{linenomath*}
\begin{equation}
c_{\rm 200}(M) = 5\left(\frac{M}{10^{14}h^{-1}M_{\odot}}\right)^{-0.1}
\label{cvir}
\end{equation}
\end{linenomath*} 
and the halo radius to be defined by
\begin{linenomath*} 
\begin{equation}
R_{\rm 200} = \left(\frac{3 M}{4 \pi 200 \rho_{\rm M}} \right)^{\frac{1}{3}}.
\label{Rvir}
\end{equation}
\end{linenomath*}
Together with an appropriate halo occupation distribution (HOD) model, equations \ref{DelSigDef}-\ref{Rvir} allow for the theoretical computation of $\Delta \Sigma(r_p)$ and $\gamma_t(r_p)$. We will specify the HOD models that we use in this work below, in Section \ref{sec:obsscen}. 

\subsection{Galaxy-galaxy lensing and intrinsic alignments}
\label{subsec:ggl_and_ia}
\noindent
Intrinsic alignment contributions to galaxy-galaxy lensing signals arise due to correlations between shapes of source galaxies and positions of lens galaxies which are due not to lensing but to tidal gravitational effects. 

Consider a generic cross-correlation of the shear of source galaxies with the positions of lens galaxies $\langle \gamma \delta_l \rangle$, where we have used $\gamma$ to represent shear and $\delta_l$ to represent lens galaxy overdensity. The intrinsic alignment contamination to this cross-correlation can be expressed via
\begin{linenomath*}
\begin{equation}
\langle \gamma \delta_l \rangle = \langle \gamma_{\rm L} \delta_l \rangle + \langle \gamma_{\rm IA} \delta_l \rangle
\label{GGL_IA},
\end{equation}
\end{linenomath*}
where we use $\gamma_{\rm L}$ to represent the true shear due to lensing, and $\gamma_{\rm IA}$ to represent the effective contribution to the shear due to IA.

The term $\langle \gamma_{\rm IA} \delta_l \rangle$ is generally non-zero in the case when galaxies from the source sample are in physical proximity to lens galaxies, and therefore have shapes which are correlated with the lens-galaxy positions via tidal gravitational effects. Given perfect redshift measurements for both source and lens galaxies, it would be possible to minimise or eliminate this effect by down-weighting or cutting lens-source pairs which are close along the line of sight. However, due to the large number of source galaxies employed in weak lensing measurements, in general source galaxy redshifts are determined via photometry and hence have some non-negligible uncertainty.

The general cross-correlation $\langle \gamma \delta_l \rangle$ can be taken to represent $\tilde{\gamma}_t(r_p)$ or $\widetilde{\Delta \Sigma}(r_p)$, and the contamination to either of these signals can be quantified via $\bar{\gamma}_{\rm IA}(r_p)$: the contribution to the measured tangential shear from IA, per contributing lens-source pair. For the forecasts which we undertake below, we will require fiducial predictions for this quantity, which are obtained via
\begin{linenomath*}
\begin{align}
\bar{\gamma}_{\rm IA}(r_p) &\approx \frac{w_{l+}(r_p)}{w_{ls}(r_p)+ 2 \Pi_{\rm close}},
\label{gammaIA_the}
\end{align} 
\end{linenomath*}
where $w_{ls}(r_p)$ is the projected correlation function of the positions of lens galaxies with those of source galaxies, $w_{l+}(r_p)$ is the projected cross-correlation function between lens galaxy positions and source galaxy shapes, and $\Pi_{\rm close}$, in analogy to $\Pi$ above, is the comoving line-of-sight separation within which lens-source pairs are sufficiently close along the line of sight to be affected by IA. A similar equation for $\bar{\gamma}_{\rm IA}(r_p)$ is given in, e.g., B2012, the difference being the additive factor of $2\Pi_{\rm close}$ included here in the denominator. This factor is required because the method we introduce here will assume that all physically associated lens-source pairs may be subject to IA (rather than excess pairs only), with $\Pi_{\rm close}$ being the line-of-sight separation within which pairs are expected to be physically associated. The denominator of equation \ref{gammaIA_the} can be thought of as being obtained by integrating along the line-of-sight not just over $\xi_{ls}(r_p)$, but over $\xi_{ls}(r_p)+1$. We will take $\Pi_{\rm close}$ to be $100\, {\rm Mpc/h}$, although this value is not well-determined and represents an uncertainty in our modelling of the fiducial signal. We emphasise that the choice of a $\Pi_{\rm close}$ value is required only to provide a theoretical $\bar{\gamma}_{\rm IA}(r_p)$ signal for forecasting; it does not need to be specified {\it a priori} when making a measurement.

Non-zero intrinsic alignment signals have been observed in red galaxy populations, but not yet conclusively measured in blue galaxy populations (see, e.g., \citealt{Mandelbaum2011}, \citealt{Kirk2015}); we therefore expand equation \ref{gammaIA_the} as
\begin{linenomath*}
\begin{align}
\bar{\gamma}_{\rm IA}(r_p) &\approx \frac{f_{\rm red} w_{l+}^{\rm red}(r_p) + f_{\rm blue} w_{l+}^{\rm blue}(r_p)}{f_{\rm red}w_{ls}^{\rm red}(r_p)+f_{\rm blue} w_{ls}^{\rm blue}(r_p)+ 2 \Pi_{\rm close}} \nonumber \\ &\approx \frac{f_{\rm red} w_{l+}^{\rm red}(r_p)}{w_{ls}(r_p)+ 2 \Pi_{\rm close}},
\label{gammaIA_the_fred}
\end{align} 
\end{linenomath*}
where $f_{\rm red}$ and $f_{\rm blue}$ respectively represent the fraction of red and blue source galaxies amongst those which are sufficiently close in line-of-sight separation to the lens sample to be subject to intrinsic alignments. The second line of equation \ref{gammaIA_the_fred} comes from the assumption that $w_{l+}^{\rm blue}(r_p)\approx 0$ (i.e., blue galaxies are not subject to IA at a significant level) and that blue and red galaxies cluster in the same way (which is not strictly correct  but is a sufficient approximation for the purposes of our work).

The two-halo terms of these projected correlation functions are computed via (see, for example, \citealt{Singh2014}, where we have neglected redshift-space distortions):
\begin{linenomath*}
\begin{align}
w_{ls}^{2h}(r_p) &= \frac{b_s b_l}{\pi^2} \int dz W(z) \int dk_z  \int dk_\perp  \frac{k_\perp}{k_z} P_{\delta}\left(\sqrt{k_\perp^2 + k_z^2}, z\right) \nonumber \\ &\times \sin(k_z \Pi_{\rm max}) J_0(r_p k_\perp) \label{wls_2h}  \\
w_{l+}^{2h}(r_p) &= \frac{A_I b_l C_1 \rho_c \Omega_M}{\pi^2} \int dz \frac{W(z)}{D(z)}\int dk_z  \int dk_\perp  \frac{k_\perp^3 }{(k_z^2 + k_\perp^2)k_z} \nonumber \\ &\times P_{\delta}\left(\sqrt{k_\perp^2 + k_z^2}, z\right) \sin(k_z \Pi_{\rm max}) J_2(r_p k_\perp)
\label{wlp_2h}
\end{align}
\end{linenomath*}
where $b_s$ is the large-scale bias of the source galaxies, $b_l$ is the same for the lens galaxies, and $P_\delta$ is the nonlinear matter power spectrum. We set $b_s$ and $b_l$ to their ensemble average values as computed from an appropriately chosen HOD (the HODs used in this work for different observational scenarios are discussed in Section \ref{sec:obsscen}).

Equation \ref{wlp_2h} assumes the nonlinear alignment model for IA. Here, $A_I$ controls the amplitude of IA on scales where the two-halo term dominates, and $C_1$ is a normalisation constant. Throughout this work we follow e.g. \cite{Joachimi2011} and set $C_1 \rho_c = 0.0134$, obtained via fitting the linear alignment model to low-redshift SuperCOSMOS observations. $W(z)$ is the combined window function of source and lens galaxy samples, given by (see for example \citealt{Singh2014}):
\begin{linenomath*}
\begin{equation}
W(z) = \frac{\frac{dN(z)}{dz_l}\frac{dN(z)}{dz_s}\frac{1}{\chi^{2}(z)} \left( \frac{d\chi}{dz}\right)^{-1}}{ \left(\int \, dz \frac{dN(z)}{dz_l}\frac{dN(z)}{dz_s}\frac{1}{\chi^{2}(z)} \left( \frac{d\chi}{dz}\right)^{-1}\right)}
\label{window}
\end{equation}
\end{linenomath*}
where $\frac{dN}{dz_l}$ and $\frac{dN}{dz_s}$ are the redshift distributions of the lens and source galaxies respectively.

The one-halo term of $w_{ls}(r_p)$ is computed using the standard halo model in combination with the relevant chosen HOD (again, given below in Section \ref{sec:obsscen}). The NFW profile is assumed, and concentration and virial radius are given by equation \ref{cvir} and \ref{Rvir} respectively. 

$w_{l+}^{1h}(r_p)$ is calculated using the halo model for IA as introduced in \cite{Schneider2010}. The relevant one-halo power spectrum is:
\begin{linenomath*}
\begin{equation}
P^{1h}_{l+}(k,z) = a_h \frac{(k/p_1)^2}{1+ (k/p_2)^{p_3}}
\label{P1hIA}
\end{equation}
\end{linenomath*}
where
\begin{linenomath*}
\begin{equation}
p_i = q_{i1} {\rm exp}(q_{i2} z^{q_{i3}}).
\label{pi}
\end{equation}
\end{linenomath*}
The parameters $q_{ij}$ are fit in \cite{Schneider2010}. In \cite{Singh2014}, the $q_{i1}$ parameters are adjusted to better fit the BOSS LOWZ galaxy sample; here we assume the $q_{i1}$ parameters of \cite{Singh2014} and all other $q_{ij}$ parameters from \cite{Schneider2010}. $w_{l+}^{1h}(r_p)$ can then be found via
\begin{linenomath*}
\begin{equation}
w_{l+}^{1h}(r_p) = \int \frac{dk_\perp}{k_\perp}{2\pi} P^{1h}_{l+}(k_\perp,z) J_0(r_p k_\perp).
\label{wg1h}
\end{equation}
\end{linenomath*}
Given the capability to theoretically calculate the one- and two-halo terms of both $w_{ls}(r_p)$ and $w_{l+}(r_p)$, theoretical fiducial values for $\bar{\gamma}_{\rm IA}(r_p)$ can be computed using equation \ref{gammaIA_the}. Note that we do not incorporate halo-exclusion terms in our fiducial calculations of $w_{l+}(r_p)$ and $w_{ls}(r_p)$, but instead simply add one-halo and two-halo terms. The choice to simplify our calculations in this way has a minor effect on the shape of the fiducial signals calculated, however the magnitude of this effect is negligible relative to the variation in signal-to-noise within the IA-measurement scenarios explored below.

\subsection{Existing methods for measuring intrinsic alignments}
\label{subsec:existing_methods}
\noindent
Several methods exist in the literature to measure or constrain the intrinsic alignment contribution to the galaxy-galaxy lensing signal. Because lensing measurements typically rely on source galaxies with redshifts determined photometrically, it is useful that procedures for mitigating the effect of IA take this into account. In several existing methods, source galaxies are separated into two or more bins using photometric redshifts, and the lensing and IA signals are estimated simultaneously via the assumption that the source galaxy sample(s) which are more separated in redshift from the lens galaxies will contain fewer galaxies which are subject to IA. 

In \cite{Hirata2004b} the intrinsic alignment contribution to an SDSS galaxy-galaxy lensing measurement was constrained, under the assumption that a source sample with higher photometric redshifts had no intrinsic alignment contribution. Methods by B2012 and \cite{Chisari2014} later relaxed this assumption, while still relying on the idea that fewer galaxies in higher photometric-redshift source samples would be subject to IA. In a similar vein, \cite{Joachimi2008} and \cite{Joachimi2010} proposed methods to null or boost the intrinsic alignment signal in cosmic shear measurements using its characteristic redshift dependence. In order to demonstrate the utility of the method which we present in this work, we will compare against the method put forward in B2012. We now describe this method in detail.

In the methodology of B2012, it is assumed that source galaxy redshifts are photometric while lens galaxy redshifts are spectroscopic. Two measurements of the galaxy-galaxy lensing quantity $\widetilde{\Delta \Sigma}$ are then considered. The first, which we label $a$, is for a source sample defined by the requirement that for a given lens redshift $z_l$, the source (photometric) redshift $z_{ph}$ satisfies $z_l < z_{ph}< z_l + \Delta z$, where $\Delta z$ is chosen to jointly optimise signal-to-noise of the lensing measurement and intrinsic alignment constraint on a per-survey basis. The second sample, called $b$, is chosen in a complementary way such that $z_{ph} > z_{l} + \Delta z$. Given these two measurements of $\widetilde{\Delta\Sigma}$, it is possible to solve both for the lensing signal and for the contribution to the tangential shear due to IA. The expression for the latter is given by B2012:
\begin{linenomath*}
\begin{equation}
\bar{\gamma}^{\rm IA} = \frac{c_z^{(a)} \widetilde{\Delta\Sigma}_a - c_z^{(b)} \widetilde{\Delta\Sigma}_b}{(B_a-1)c_z^{(a)} \langle \tilde{\Sigma}_c\rangle_{\rm ex}^{(a)} -(B_b-1)c_z^{(b)} \langle \tilde{\Sigma}_c\rangle_{\rm ex}^{(b)}}.
\label{gammaIA_Blazek}
\end{equation}
\end{linenomath*}
In the form of this method originally introduced in B2012, it is assumed that only excess lens-source pairs are subject to IA (that is, only pairs which statistically contribute to a positive correlation between lenses and sources). $c_z^{(i)}$ is given by $\left(1 + b_z^{(i)}\right)^{-1}$ where $b_z^{(i)}$ is the photometric-redshift bias to $\widetilde{\Delta\Sigma}_i$. $B_i$ is the boost factor, which quantifies the presence of excess galaxies in sample $i$, and $\langle \tilde{\Sigma}_c\rangle_{\rm ex}^{(i)}$ is the average critical surface density for excess galaxies. Dependence on $r_p$ has been omitted in equation \ref{gammaIA_Blazek} for clarity. 

$c_z$, $B$, and $\langle \tilde{\Sigma}_c\rangle_{\rm ex}$ are explicitly given for a generic source sample by:
\begin{linenomath*}
\begin{align}
&c_z^{-1} = b_z+1 = \frac{B(r_p) \sum\limits^{\rm lens}_{j} \tilde{w}_j \tilde{\Sigma}_{c,j} \Sigma_{c,j}^{-1}}{\sum\limits^{\rm lens}_{j} \tilde{w}_j} \approx \frac{\sum\limits^{\rm rand}_{j} \tilde{w}_j \tilde{\Sigma}_{c,j} \Sigma_{c,j}^{-1}}{\sum\limits^{\rm rand}_{j} \tilde{w}_j} \label{photozbias} \\
&B(r_p) = \frac{\sum\limits_{j}^{\rm lens}\tilde{w}_j}{\sum\limits_{j}^{\rm rand} \tilde{w}_j} \label{boost1} \\
&\langle \tilde{\Sigma}_c\rangle_{\rm ex} (r_p) =  \frac{\sum\limits_{j}^{\rm excess} \tilde{w}_j \tilde{\Sigma}_{c,j}}{\sum\limits_{j}^{\rm excess} \tilde{w}_j} = \frac{\sum\limits_{j}^{\rm lens} \tilde{w}_j \tilde{\Sigma}_{c,j}- \sum\limits_{j}^{\rm rand} \tilde{w}_j \tilde{\Sigma}_{c,j}}{\sum\limits_{j}^{\rm lens} \tilde{w}_j - \sum\limits_{j}^{\rm rand} \tilde{w}_j},\label{Sigmaex}
\end{align}
\end{linenomath*}
where in equation \ref{photozbias}, the second equality makes use of the fact that the true value of  $\Sigma_c^{-1}$ is approximately zero for source galaxies at or very near the lens redshift. Sums in equations \ref{photozbias}, \ref{boost1}, and \ref{Sigmaex}, and similar sums below, should be interpreted as being over lens-source pairs with source galaxies in the relevant sample. A label of `lens' indicates a sum over all such pairs, while a label of `excess' implies a sum over only those pairs which are statistically in excess of what would be present in the case without clustering. A label of `rand' indicates that the sum should be performed over a set of `lenses' which are sampled randomly from the same redshift distribution and window function as the true set of lenses. 

We note particularly that equation \ref{photozbias} for $c_z$ requires a sum over the product of $\tilde{\Sigma}_{c,j}$, the estimated critical surface density (using photometric redshifts for source galaxies) and $\Sigma_{c,j}^{-1}$, the inverse of the true critical surface density (using spectroscopic redshifts for source galaxies). This sum is in practice therefore taken over only the subsample of lens-source pairs for which sources have spectroscopic redshifts. We assume that in the case in which a representative spectroscopic subsample of sources is unavailable, a re-weighting method (see, for example, \citealt{Lima2008}) is used to approximate a representative subsample as closely as possible. Such a method will be imperfect when the spectroscopic subsample completely neglects parts of the source galaxy parameter space (in terms of e.g. colour or magnitude), or if the rate of redshift failure at a fixed point in colour-magnitude space depends on redshift. Biases can similarly arise if the selection function of the overlapping spectroscopic survey (for example, in colour space) cannot easily be reproduced by the wide-field survey which images the sources. In such cases, $c_z$ will be subject to a systematic uncertainty. 

As stated, the objective of the B2012 method to which we will compare is to solve simultaneously for both $\bar{\gamma}^{\rm IA}(r_p)$ and $\Delta \Sigma(r_p)$. Therefore, the weights, $\tilde{w}_j$, are chosen in that work to optimise the signal-to-noise of $\Delta \Sigma(r_p)$:
\begin{linenomath*}
\begin{equation}
\tilde{w}_j = \frac{1}{\tilde{\Sigma}_{c,j}^2 (\sigma_\gamma^2 + (\sigma_e^j)^2)},
\label{weights_Blazek}
\end{equation}
\end{linenomath*}
where $\sigma_\gamma$ is the contribution due to shape-noise and $\sigma_e^j$ is the measurement error associated with the source of lens-source pair $j$. This choice of weights is sub-optimal for the estimation of IA, due in part to the downweighting of nearby lens-source pairs. We do not advocate its use with our proposed method, and will introduce a different weighting scheme below for this purpose. We nevertheless use the weights of equation \ref{weights_Blazek} when making calculations in the B2012 method to enable comparison with the measurements of that work. One might then ask whether comparisons made between our method, with a more-optimal weighting scheme, and the B2012 method, with this less-optimal scheme, are meaningful. To address this issue, we have checked the effect of using a modified version of the B2012 method which is formulated in terms of tangential shear and uses the redshift-independent weights introduced below in equation \ref{weights}. We find no qualitative changes to our results in this scenario, and the quantitative changes which do occur do not affect our overall conclusions (for example, the ratio of signal-to-noise quantities presented in Figure \ref{fig:StoNstat} below would be reduced by 30\% in the LSST+DESI observational scenario). We mention a further implication of this alternative weighting scheme for the B2012 method in Section \ref{subsec:statresults}.

\section{Measuring intrinsic alignments with multiple shear estimates}
\label{sec:newmethod}
\noindent
Having now provided theoretical background, we introduce our new method to measure the scale dependence of the intrinsic alignment contribution to the galaxy-galaxy lensing signal.

We consider two estimates of the tangential shear obtained via different methods but using the same source and lens galaxy samples, which we will call $\tilde{\gamma}_t(r_p)$ and $\tilde{\gamma}_t^\prime(r_p)$. These are given as:
\begin{linenomath*}
\begin{align}
\tilde{\gamma}_t(r_p) &= B(r_p) \frac{\sum\limits_{j}^{\rm lens} \tilde{w}_j \tilde{\gamma}_j}{\sum\limits_{j}^{\rm lens} \tilde{w}_j} = B(r_p) m \left( \frac{\sum\limits_{j}^{\rm lens} \tilde{w}_j \gamma_{\rm L}^j}{\sum\limits_{j}^{\rm lens} \tilde{w}_j}+ \frac{\sum\limits_{j}^{\rm lens} \tilde{w}_j \gamma_{\rm IA}^j}{\sum\limits_{j}^{\rm lens} \tilde{w}_j} \right) \label{gammaus} 
\end{align}
\end{linenomath*}
and
\begin{linenomath*}
\begin{align} 
\tilde{\gamma}_t^\prime (r_p) &= B(r_p) \frac{\sum\limits_{j}^{\rm lens} \tilde{w}_j \tilde{\gamma}_j^\prime}{\sum\limits_{j}^{\rm lens} \tilde{w}_j} = B(r_p)m^\prime\left(\frac{\sum\limits_{j}^{\rm lens} \tilde{w}_j \gamma_{\rm L}^j}{\sum\limits_{j}^{\rm lens} \tilde{w}_j}+\frac{a \sum\limits_{j}^{\rm lens} \tilde{w}_j \gamma_{\rm IA}^j}{\sum\limits_{j}^{\rm lens} \tilde{w}_j}\right) \label{gammaprimeus}
\end{align}
\end{linenomath*}
where all sums are over lens-source pairs, $\gamma_L^j$ is the tangential shear due to lensing for a given lens-source pair $j$, and $\gamma_{\rm IA}^j$ is the contribution to the shear signal due to IA for lens-source pair $j$. $B(r_p)$ is once again the boost factor, which is included to ensure that the tangential shear is normalised in the standard way. $m = (1+\delta m)$ and $m^\prime = (1+\delta m^\prime)$ are, for each shear estimate, the residual multiplicative bias remaining after calibration. $a$ is the constant by which the measured IA amplitudes are offset one from the other. For reference, as we define $a$ here, \cite{Singh2016} find $a\approx0.7-0.8$ for the three method pairs formed by de Vaucouleurs shapes, isophotal shapes, and re-Gaussianization. These pairings of shear estimates are not expected to be optimal; we provide their $a$ values simply for reference and would expect more optimal methods to yield smaller $a$ values (as discussed below). Note that our definition of $a$ is slightly different from the ratio of IA amplitude values used to describe this effect in \cite{Singh2016}.

In contrast to the method described in Section \ref{subsec:existing_methods}, our objective is not to simultaneously estimate both $\Delta \Sigma (r_p)$ and $\bar{\gamma}_{\rm IA}$, but simply to measure the scale dependence of $\bar{\gamma}_{\rm IA}$. We therefore choose weights differently than in the above case:
\begin{linenomath*}
\begin{equation}
\tilde{w}_j = \frac{1}{\sigma_\gamma^2 + (\sigma_e^j)^2}.
\label{weights}
\end{equation}
\end{linenomath*}
This is a typical choice of weights in measurements of lensing tangential shear. While not being explicitly designed to optimise for $\bar{\gamma}_{\rm IA}(r_p)$, this choice is more optimal than e.g. the weights of equation \ref{weights_Blazek} because it dispenses with the factor of $\tilde{\Sigma}_{c}^{-2}$, which down-weights the very lens-source pairs expected to be subject to IA. It is nevertheless likely that there exist more optimal weight choices than equation \ref{weights}, and therefore the forecasts conducted below may not reflect the highest possible signal-to-noise available via the proposed method.

The success of our proposed method is dependent upon using the same weighting scheme when computing $\tilde{\gamma}_t(r_p)$ and $\tilde{\gamma}^\prime_t(r_p)$. Therefore, if the two shear estimates in question result in different values of $\sigma_\gamma$, or different per-galaxy values of $\sigma_e$, it is necessary to use a version of equation \ref{weights} which adopts the same value of these for both methods (e.g. by adopting their average value). This ensures the required cancellation when taking the difference of $\tilde{\gamma}_t(r_p)$ and $\tilde{\gamma}^\prime_t(r_p)$, as we do now.

Subtracting equation \ref{gammaprimeus} from \ref{gammaus},  we obtain:
\begin{linenomath*}
\begin{align}
\tilde{\gamma}_t(r_p) - \tilde{\gamma}_t^\prime(r_p) &= (1-a) B(r_p) \frac{\sum\limits_{j}^{\rm lens} \tilde{w}_j \gamma_{\rm IA}^j}{\sum\limits_{j}^{\rm lens} \tilde{w}_j} = (1-a) \frac{\sum\limits_{j}^{\rm lens} \tilde{w}_j \gamma_{\rm IA}^j}{\sum\limits_{j}^{\rm rand} \tilde{w}_j}.
\label{subtract_gammat}
\end{align}
\end{linenomath*}

Terms involving $m$ and $m^\prime$ are no longer present in equation \ref{subtract_gammat}. We have assumed that calibration is performed via a method in which any such residual uncertainty ($\delta m$ or $\delta m^\prime$ above) is demonstrably sub-dominant (see, for example, \citealt{Bernstein2016, Sheldon2017}). If this is not the case, two additional terms would be added to the right-hand-side of equation \ref{subtract_gammat}, representing an additive bias: one proportional to $(m-m^\prime)\gamma_{\rm L}$ and one proportional to $(m-am^\prime)\bar{\gamma}_{\rm IA}$. We estimate that in order to limit this additive bias to at most $n$-percent the value of $(1-a)\bar{\gamma}_{\rm IA}$, for a galaxy-galaxy lensing measurement involving sources from LSST and lenses from the Dark Energy Spectroscopic Instrument (see Section \ref{sec:obsscen} below for details on this observational scenario), we would require that the individual absolute values of $\delta m$ and $\delta m^\prime$ not exceed $ \approx n \% \times (1-a)$.

We have also assumed in equation \ref{subtract_gammat} that any residual additive biases to $\tilde{\gamma}_t(r_p)$ and $\tilde{\gamma}_t^\prime(r_p)$ exactly cancel, which need not be the case (see, for example, \citealt{Sheldon2004}). To circumvent this potential issue, estimators for tangential shear which incorporate subtraction of the signal measured about random points from that measured about lenses could be used, which would render additive biases to each of $\tilde{\gamma}_t(r_p)$ and $\tilde{\gamma}_t^\prime(r_p)$ individually negligible.

We consider a sample of lens-source pairs which is defined by the requirement that for a given lens-source pair, lens redshift and source photometric redshift are sufficiently close that we would naively expect them to be physically associated. As mentioned above, we will take this maximum line-of-sight separation to be $100$ $ {\rm Mpc} / {\rm h}$ in this work. The quantity which we aim to measure, up to a constant factor, is  $\bar{\gamma}_{\rm IA}$: the tangential shear due to intrinsic alignments per contributing lens-source pair. To obtain this quantity from equation \ref{subtract_gammat}, we must multiply both sides by an additional factor, which ensures that we divide by the sum of weights of contributing pairs only:
\begin{linenomath*}
\begin{align}
\frac{\sum\limits_{j}^{\rm rand} \tilde{w}_j} {\sum\limits_{j}^{\rm excess} \tilde{w}_j + \sum\limits_{j}^{\rm rand,\,close} \tilde{w}_j} 
\equiv \frac{1}{B(r_p) - 1 + F},
\label{factor_renom}
\end{align}
\end{linenomath*}
where we have defined $F$:
\begin{linenomath*}
\begin{equation}
F = \frac{\sum\limits_{j}^{\rm rand,\,close} \tilde{w}_j} {\sum\limits_{j}^{\rm rand} \tilde{w}_j}
\label{F_def}
\end{equation}
\end{linenomath*}
and the label `rand, close' indicates that the sum should be taken over pairs in which source and lens are sufficiently close as measured by spectroscopic redshift that we would expect them to be intrinsically aligned, and the `lenses' are drawn randomly from the lens redshift distribution and window function. 

$F$ can evidently not be computed on a pair-by-pair basis, as we do not expect to have access to the spectroscopic redshift of all source galaxies. Instead it should be computed statistically via $\frac{dN}{dz_s}$ and $p(z_s, z_{ph})$, where the former is the number density of sources with respect to spectroscopic redshift, and the latter is the probability of a source galaxy with measured photometric redshift $z_{ph}$ having true (spectroscopic) redshift $z_s$, or vice-versa. In this way, $F$ is given by:
\begin{linenomath*}
\begin{align}
F = \frac{\int dz_l \frac{dN}{dz_l}\int dz_{ph} \tilde{w}(z_{ph}) \int_{z_-(z_l)}^{z_+(z_l)} dz_s\frac{dN}{dz_s} p(z_s, z_{ph})}{\int dz_l \frac{dN}{dz_l}\int dz_{ph}  \tilde{w}(z_{ph}) \int dz_s\frac{dN}{dz_s} p(z_s, z_{ph})}
\label{ncorr_2}
\end{align}
\end{linenomath*}
where $z_{+}(z_l)$ and $z_{-}(z_l)$ are, respectively, the maximum and minimum spectroscopic redshifts at which we would expect source galaxies to be intrinsically aligned, for a given lens redshift $z_l$. $\frac{dN}{dz_l}$ is the number density of lens galaxies with respect to (spectroscopic) redshift, and integrals without explicit limits should be taken as being over the integration variable's full range. 

Equation \ref{ncorr_2} makes explicit the fact that because $F$ encodes information about true source galaxy redshifts, it must be calculated using $\frac{dN}{dz_s}$ and $p(z_s, z_{ph})$ as estimated from the spectroscopic subsample of sources. Much like $c_z$ in the B2012 method above, in the case of an inadequately representative spectroscopic subsample or imperfect re-weighting, $F$ may therefore become a source of systematic error. 

We finally multiply equation \ref{subtract_gammat} by the factor defined in equation \ref{factor_renom} to get
\begin{linenomath*}
\begin{equation}
(1-a) \bar{\gamma}_{\rm IA}(r_p) = \frac{\tilde{\gamma}_t(r_p) - \tilde{\gamma}_t^\prime(r_p)}{B(r_p)-1 + F}.
\label{IAsol3}
\end{equation}
\end{linenomath*}
Equation \ref{IAsol3} is the fundamental expression of our method. It allows us to measure the intrinsic alignment contribution to galaxy-galaxy lensing up to a poorly-known constant in order to gain information about its scale dependence. We now prepare to test how well this method is expected to perform compared to the existing method described in Section \ref{subsec:existing_methods}.

\section{Observational scenarios for forecasting}
\label{sec:obsscen}
\noindent
To evaluate the effectiveness of our proposed method for measuring the scale dependence of $\bar{\gamma}_{\rm IA}$, we select two observational scenarios in which to forecast expected performance. 

The first of these, which we will call the `SDSS' case, assumes that lens and source galaxies are both drawn from Sloan Digital Sky Survey (SDSS) data. For the other, which we will call the `LSST+DESI' case, we consider a scenario which combines data from two upcoming surveys, with sources from the Large Synoptic Survey Telescope (LSST) and lenses from the Dark Energy Spectroscopic Instrument Luminous Red Galaxy sample (DESI LRGs). These two choices ensure that we can both compare our predictions to the actual measurement of $\bar{\gamma}_{\rm IA}$ by B2012 (in the SDSS case), and explore how our proposed method may perform for a next-generation measurement (in the LSST+DESI scenario). 

In the SDSS case, lens galaxies are assumed to be from the SDSS LRG sample as selected in B2012 (see also \citealt{Kazin2010} and their `DR7-Dim' sample), with a surface density of $n_l = 8.7/{\rm deg}^2$ (\citealt{Mandelbaum2013}). The median redshift of this sample is $z=0.28$ (\citealt{Kazin2010}). Source galaxies are assumed to be from the sample described in \cite{Reyes2012}, with an effective surface density of $n_{\rm eff} = 1/{\rm arcmin}^2$. The per-component rms distortion of the source sample is $\epsilon_{\rm rms} = 0.36$; with responsivity $\mathcal{R}=(1 - \epsilon_{\rm rms}^2) \approx 0.87$, this results in $\sigma_\gamma = \epsilon_{\rm rms} / (2 \mathcal{R}) = 0.21$ (\citealt{Reyes2012}). The overlapping sky area of these lens and source samples is taken as 7131 ${\rm deg}^2$ (B2012).

The LSST+DESI observational scenario assumes lens galaxies from the anticipated DESI sample of LRGs, which we assume to have a surface density of $300/{\rm deg}^2$ \citep{DESIExperiment}. The effective redshift of the sample is taken as $z=0.77$ (estimated from Figure 3.8 of \citealt{DESIExperiment}). Source galaxies in this scenario are taken to be from the final LSST lensing sample, with an expected effective surface density of $n_{\rm eff} = 26/{\rm arcmin}^2$, and a per-component rms ellipticity of $\epsilon_{\rm rms} = 0.26$, which with the ellipticity definition of \cite{Chang2013} is equivalent to $\sigma_\gamma=0.26$. These source and lens samples would be expected to have an overlapping sky area of 3000 ${\rm deg}^2$ \citep{Schmidt2014}. 

\begin{figure*}
\centering
\subfigure{\includegraphics[width=0.35\textwidth]{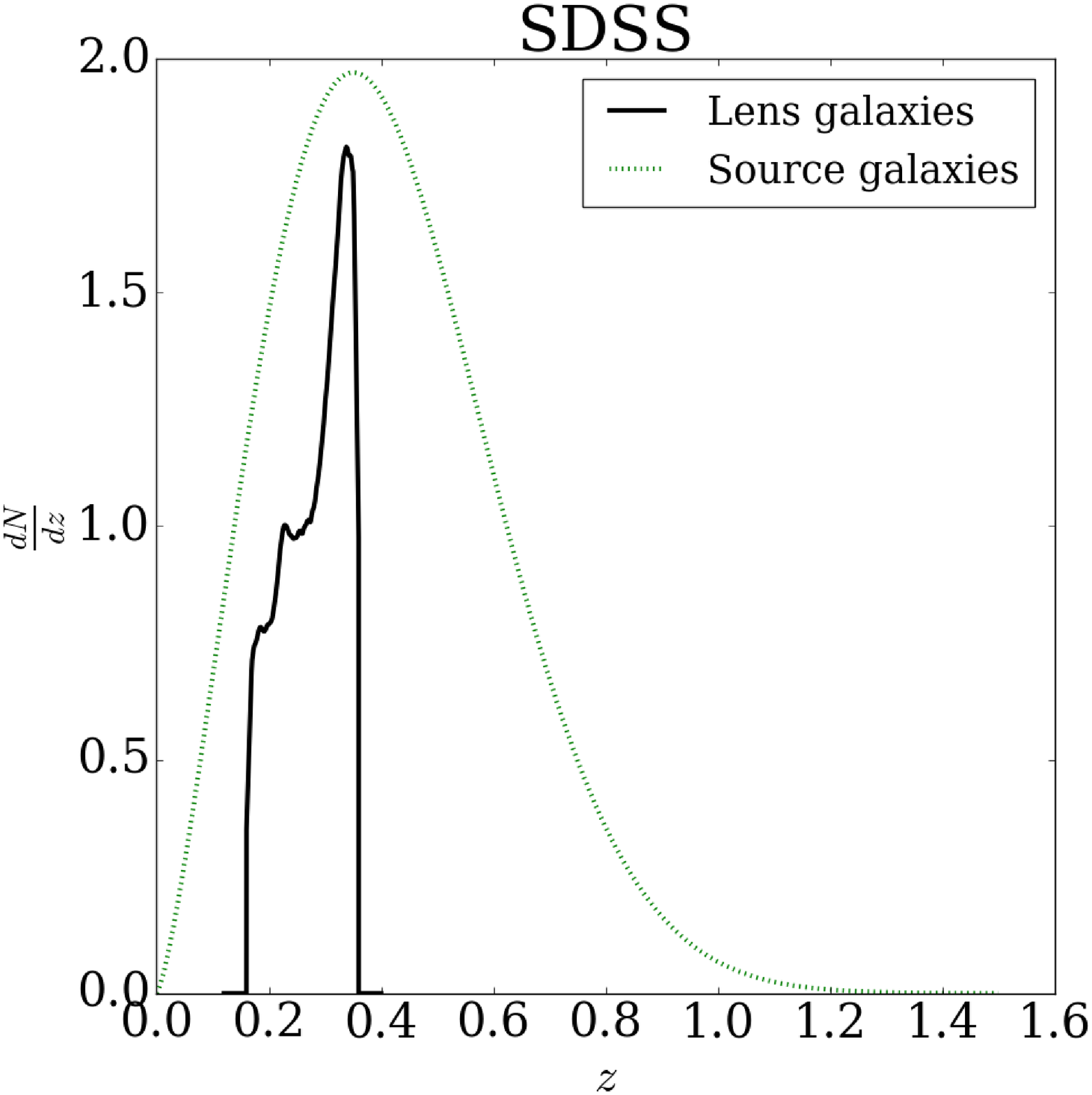}}
\subfigure{\includegraphics[width=0.35\textwidth]{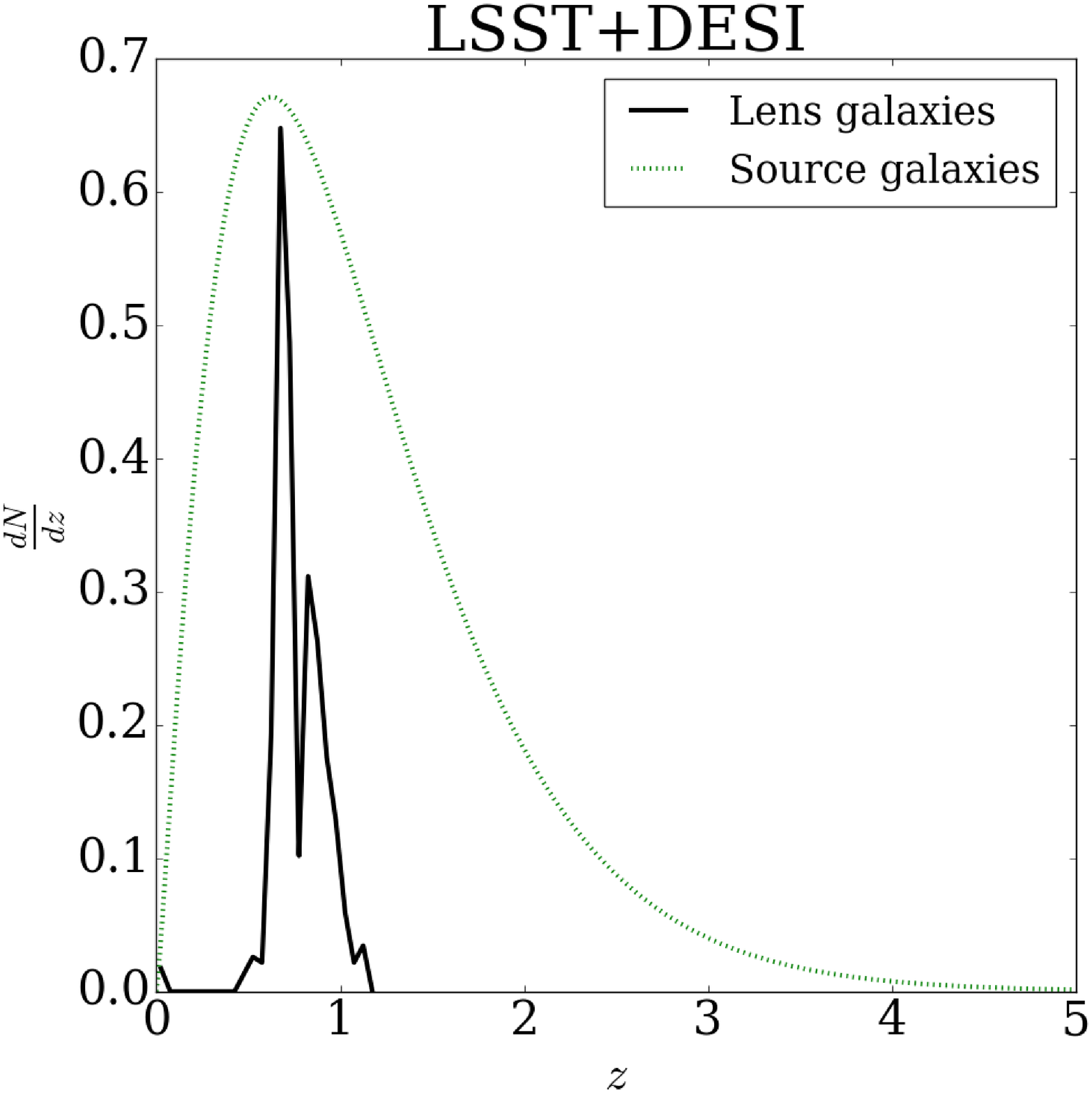}}
\caption{The assumed spectroscopic redshift distribution of lens and source galaxies for both observational scenarios (left: SDSS, right: LSST+DESI). Source distributions are shown normalised individually to unity over the redshift range, while lens distributions are arbitrarily scaled (for visualisation purposes only).}
\label{fig:dNdz}
\end{figure*}

Further assumptions and details associated with each of these observational scenarios are as follows:

\paragraph*{Redshift distribution of lenses:} The number density of the SDSS LRG sample is shown as a function of redshift in Figure 2 of \cite{Kazin2010}. We smooth with a box filter and perform a conversion to appropriate units to obtain the redshift distribution of SDSS LRGs. The expected redshift distribution of the DESI LRG sample is given in Figure 3.8 of \cite{DESIExperiment}. Both distributions are plotted in Figure \ref{fig:dNdz}.

\paragraph*{Redshift distribution of sources:}The redshift distribution (in terms of spectroscopic redshifts $z_{\rm s}$) of source galaxies is given for the SDSS source sample as:
\begin{linenomath*}
\begin{equation}
\frac{dN}{dz_{s}} \propto \left(\frac{z_{s}}{z_*}\right)^{\alpha-1} {\rm exp}\left(-\frac{1}{2}\left(\frac{z_{s}}{z_*}\right)^2\right)
\label{dndz_sdss_source}
\end{equation}
\end{linenomath*}
where $\alpha=2.338$ and $z_*=0.303$ \citep{Nakajima2011}. For the LSST source sample, the distribution is given by:
\begin{linenomath*}
\begin{equation}
\frac{dN}{dz_{s}} \propto z_{s}^{\tilde{\alpha}} {\rm exp}\left(-\left(\frac{z_{s}}{z_0}\right)^\beta\right)
\label{dndz_lsst_sources}
\end{equation}
\end{linenomath*}
where $\tilde{\alpha}=1.24$, $z_0=0.51$ and $\beta=1.01$ \citep{Chang2013}. In both cases, $\frac{dN}{dz_{s}}$ is appropriately normalised over the redshift range once convolved with the photometric redshift model. The assumed source redshift distributions are plotted in Figure \ref{fig:dNdz}. 

\paragraph*{Model for photometric redshifts:} In both observational scenarios under consideration, source galaxy redshifts are photometrically determined. The source galaxy redshift distribution in terms of photometric redshift $z_{ph}$ is given by:
\begin{linenomath*}
\begin{equation}
\frac{dN}{dz_{ph}} = \frac{\int dz_{s} p(z_{s}, z_{ph}) \frac{dN}{dz_s}}{\int dz_{ph} \int dz_{s} p(z_{s}, z_{ph}) \frac{dN}{dz_s}}.
\label{photoz_dndz}
\end{equation}
\end{linenomath*}
We choose a simple Gaussian model for $p(z_{s}, z_{ph})$ in both observational scenarios:
\begin{linenomath*}
\begin{equation}
p(z_{s}, z_{ph}) = \frac{1}{\sqrt{2\pi}\sigma_z} {\rm exp}\left(-\frac{\left(z_{ph} - z_{s}\right)^2}{2 \sigma_z^2} \right)
\label{pzph}
\end{equation}
\end{linenomath*}
where $\sigma_z$ is taken to be $0.11(1+z_{s})$ for the SDSS source sample (B2012) and $0.05(1+z_{s})$ for the LSST source sample \citep{Chang2013}.
\paragraph*{$a_h$ and $A_I$:} $a_h$ and $A_I$ are the fiducial amplitudes of the one-halo and two-halo terms of $w_{l+}(r_p)$, respectively. A fit to a power law in luminosity for each of these quantities in provided in \cite{Singh2014}:
\begin{linenomath*}
\begin{align}
A_I(L) &= \alpha \left( \frac{L}{L_p}\right)^\beta \label{Aipowerlaw} \\
a_h(L) &= \alpha_h \left( \frac{L}{L_p}\right)^{\beta_h} \label{ahpowerlaw} 
\end{align}
\end{linenomath*}
where $L$ is the r-band luminosity and $L_p$ is the pivot luminosity corresponding to an absolute r-band magnitude of $-22$. Parameters are fit in \cite{Singh2014} using data from the SDSS BOSS LOWZ sample, and found to be $\alpha=4.9 \pm 0.6$, $\beta=1.30\pm0.27$, $\alpha_h=0.081 \pm 0.012$, and $\beta_h = 2.1 \pm 0.4$. We take the best-fit values of these parameters as their fixed values in estimating $a_h$ and $A_I$ for our fiducial calculations.

In order to then determine the appropriate values of $a_h$ and $A_I$ for the scenarios under consideration, we use a Schechter luminosity function \citep{Schechter1976} and integrate equations \ref{Aipowerlaw} and \ref{ahpowerlaw} over luminosity. Following \cite{Krause2015}, we use Schechter function parameters for red galaxies from \cite{Loveday2012} and \cite{Faber2007}. We assume the limiting $r$-band apparent magnitude of the SDSS shape sample to be $22$ (see Figure 3 of \citealt{Reyes2012}), and the same quantity for the LSST lensing sample to be $25.3$ \citep{LSSTScienceBook2}. As a result, we find $a_h= 5.6 \times 10^{-3}$ and $A_I= 0.65$ for the SDSS scenario, and $a_h=0.016$, $A_I=1.2$ for the LSST+DESI case. 

\paragraph*{Halo occupation distributions:} When computing the fiducial value of $\bar{\gamma}_{\rm IA}(r_p)$, as well as the cosmic variance terms of the covariance matrices for $\tilde{\gamma}_t$ and $\widetilde{\Delta \Sigma}$ (see Appendix \ref{app:cov}), we must specify a halo occupation distribution (HOD) for both the lens and the source samples. We use this to calculate one-halo terms of two-point functions, as well as to obtain the large-scale galaxy bias, and hence the two-halo term. For the SDSS LRGs, we use the HOD fit to this same galaxy sample in \cite{Reid2009}, which yields a large scale bias value of $b_l=2.2$. For both the SDSS and LSST source samples, we use the HOD developed in \cite{Zu2015}. This HOD has the benefit of accepting as input the number density of galaxies, allowing its use for both source samples. The galaxy bias of both source samples, according to this HOD, is $b_s=1.3$.  For the DESI LRGs, lacking a better option, we employ an HOD fit to the SDSS BOSS CMASS sample \citep{More2015} when computing one-halo terms. This is not ideal as it is fit to a different sample and does not accept number density of galaxies as input. However, as we will show below, most of the power of our proposed method is not deep within the one-halo regime, so we do not expect this sub-optimal choice to have a significant effect on our overall results. Recognising that this HOD is not a perfect choice, we do not in this case obtain a large-scale bias value from it, but rather from the expression $b(z) = 1.7 / D(z)$, given in \cite{DESIExperiment}, which at the effective redshift of the DESI LRGs results in $b_l=3.9$. Where possible, we compare our calculation of mean central and satellite galaxy occupation numbers to those computed with the Halotools package \citep{Hearin2017} and find agreement.

Note also that in computing quantities such as power spectra and correlation functions using these HODs, it is important to pair them with theoretical quantities (e.g. the halo mass function) which are calculated using appropriate values of cosmological parameters. We must use the parameter values which were either fixed in fitting HOD parameters or simultaneously fit with HOD parameters. Otherwise, the observables we calculate will not match the sample in question. Therefore, in this context only, we divert from our default cosmological parameters to use parameter values as given in the references above for each HOD, in order to best replicate observable quantities.

\paragraph*{Measurement noise, $\sigma_e$:} We take the measurement noise, $\sigma_e$, required to compute the weight of each lens-source pair as in equation \ref{weights}, to be related to the average per-source-galaxy signal-to-noise via \citep{Bernstein2002}:
\begin{linenomath*}
\begin{equation}
\sigma_e = \frac{2}{\langle S/N \rangle }.
\label{sigeSDSS}
\end{equation}
\end{linenomath*}
We take the average $S/N$ to be $15$ for the SDSS scenario. For LSST, we take $\sigma_e=0.128$ \citep{Chang2013}, equivalent to an average per-galaxy signal-to-noise of $15.6$. 

\paragraph*{Boost factors:} The boost factor, defined above in equation \ref{boost1}, is the ratio of the sum of weights over all lens-source pairs in a given sample, to the sum of weights of the same sample with randomly distributed lenses. In other words, it quantifies the degree to which correlation augments the number of galaxy pairs in the sample. To compute the boost for the various samples required for our analysis, we evaluate the expression (see for example B2012)
\begin{linenomath*}
\begin{align}
B(&r_p)-1 =\Bigg(\int dz_{l} \frac{dN}{dz_{l}} \int_{z_{-}(z_l)}^{z_{+}(z_l)} dz_{ph} \tilde{w}(z_{ph}, z_{l}) \nonumber \\ &\times \int dz_s \frac{dN}{dz_{s}} \xi_{ls}(r_p, \Pi(z_{s}); z_{l}) p(z_{s}, z_{ph})\Bigg) \nonumber \\ &\times \left(\int dz_{l} \frac{dN}{dz_{l}} \int_{z_{-}(z_l)}^{z_{+}(z_l)} dz_{ph} \tilde{w}(z_{ph}, z_{l}) \int dz_{s} \frac{dN}{dz_{s}}p(z_{s}, z_{ph})\right)^{-1}
\label{boost_analytic}
\end{align} 
\end{linenomath*}
where $z_{+}(z_l)$ and $z_{-}(z_l)$ are, respectively, the upper and lower photometric cuts which define the source sample for a given lens redshift.

\paragraph*{$\Delta z$:} For the method developed in B2012 and reviewed in Section \ref{subsec:existing_methods}, we require a value of $\Delta z$, which defines the photometric redshift range of each source sample. In the SDSS scenario, we follow B2012 and choose $\Delta z = 0.17$. In the case of LSST+DESI, we choose $\Delta z$ to optimise the signal-to-noise of $\bar{\gamma}_{\rm IA}$ in the B2012 method, and find that the optimal choice is $\Delta z = 0.1$. 

\paragraph*{Red fraction:} To obtain $f_{\rm red}$, we once again employ Schechter luminosity functions with parameters fit by \cite{Loveday2012} and \cite{Faber2007} as in \cite{Krause2015}. In this case, we consider one luminosity function with parameters fit to red galaxies only, as above, and another with parameters fit to a full sample including red and blue galaxies. To obtain the red fraction in each case, we simply take the ratio of the integrated luminosity function for red galaxies to the integrated luminosity function for all galaxies, and find the value averaged over the line-of-sight separation on which we expect pairs to be physically associated. We find $f_{\rm red}=0.27$ for SDSS, and $f_{\rm red}=0.16$ for LSST. 

\section{Results}
\label{sec:results}
\noindent
We now describe the results of forecasting constraints on the intrinsic alignment contamination to galaxy-galaxy lensing using our method, as compared to the method of B2012. We first describe an extension to the method of B2012 which will enable a fair comparison, then discuss the impact of systematic uncertainties associated with an inadequately representative spectroscopic subsample of sources. We finally present the type of measurement which may be possible in a scenario in which statistical uncertainties are dominant.

\subsection{Incorporating all physically-associated galaxies}
\label{subsec:notjustexcess}
\noindent
In order to make a fair comparison between the constraining power of our proposed method and the method of B2012, we revisit the derivation of equation \ref{gammaIA_Blazek} under the assumption that all physically associated galaxies may be subject to IA (rather than only excess galaxies). In this scenario, equation \ref{gammaIA_Blazek} becomes:
\begin{linenomath*}
\begin{equation}
\bar{\gamma}^{\rm IA}(r_p) = \frac{c_z^{(a)} \widetilde{\Delta\Sigma}_a - c_z^{(b)} \widetilde{\Delta\Sigma}_b }{(B_a-1+F_a)c_z^{(a)} \langle \tilde{\Sigma}_c\rangle_{\rm IA}^{(a)} -(B_b-1+F_b)c_z^{(b)} \langle \tilde{\Sigma}_c\rangle_{\rm IA}^{(b)}}.
\label{gammaIA_allphys}
\end{equation}
\end{linenomath*}
This result contains the new terms $F_a$ and $F_b$, where $F$ was introduced in equation \ref{F_def}, as well as the modified term $\langle \tilde{\Sigma}_c\rangle_{\rm IA}^{(i)}$, which is the average critical density over all physically associated lens-source pairs for sample $i$. $\langle \tilde{\Sigma}_c\rangle_{\rm IA}$ is given (for a generic source galaxy sample) by:
\begin{linenomath*}
\begin{align}
&\langle \tilde{\Sigma}_c\rangle_{\rm IA} (r_p) =  \frac{\sum\limits_{j}^{\rm excess} \tilde{w}_j \tilde{\Sigma}_{c,j}+ \sum\limits_{j}^{\rm rand,\,close} \tilde{w}_j \tilde{\Sigma}_{c,j}}{\sum\limits_{j}^{\rm excess} \tilde{w}_j + \sum\limits_{j}^{\rm rand,\,close} \tilde{w}_j}.\label{SigIA}
\end{align}
\end{linenomath*}
$F_i$ and $\langle \tilde{\Sigma}_c\rangle_{\rm IA}^{(i)}$ depend on sums over pairs for which the true (rather than photometric) redshift of the source galaxy is close enough to that of the lens that the pair is considered physically associated. Therefore, $F_i$ and $\langle \tilde{\Sigma}_c\rangle_{\rm IA}^{(i)}$ must be computed via integration over the source redshift distribution, and may also be subject to systematic error in the case of an inadequately representative spectroscopic subsample of sources or an imperfect re-weighting scheme. As this updated version of the B2012 method contains six terms which are in principle subject to this type of systematic uncertainty, it is reasonable to consider whether our proposed method, for which only $F$ is subject to this type of error, may be more robust to this effect. We will address this question in Section \ref{subsec:sysresults}.

\begin{figure*}
\centering
\subfigure{\includegraphics[width=0.45\textwidth]{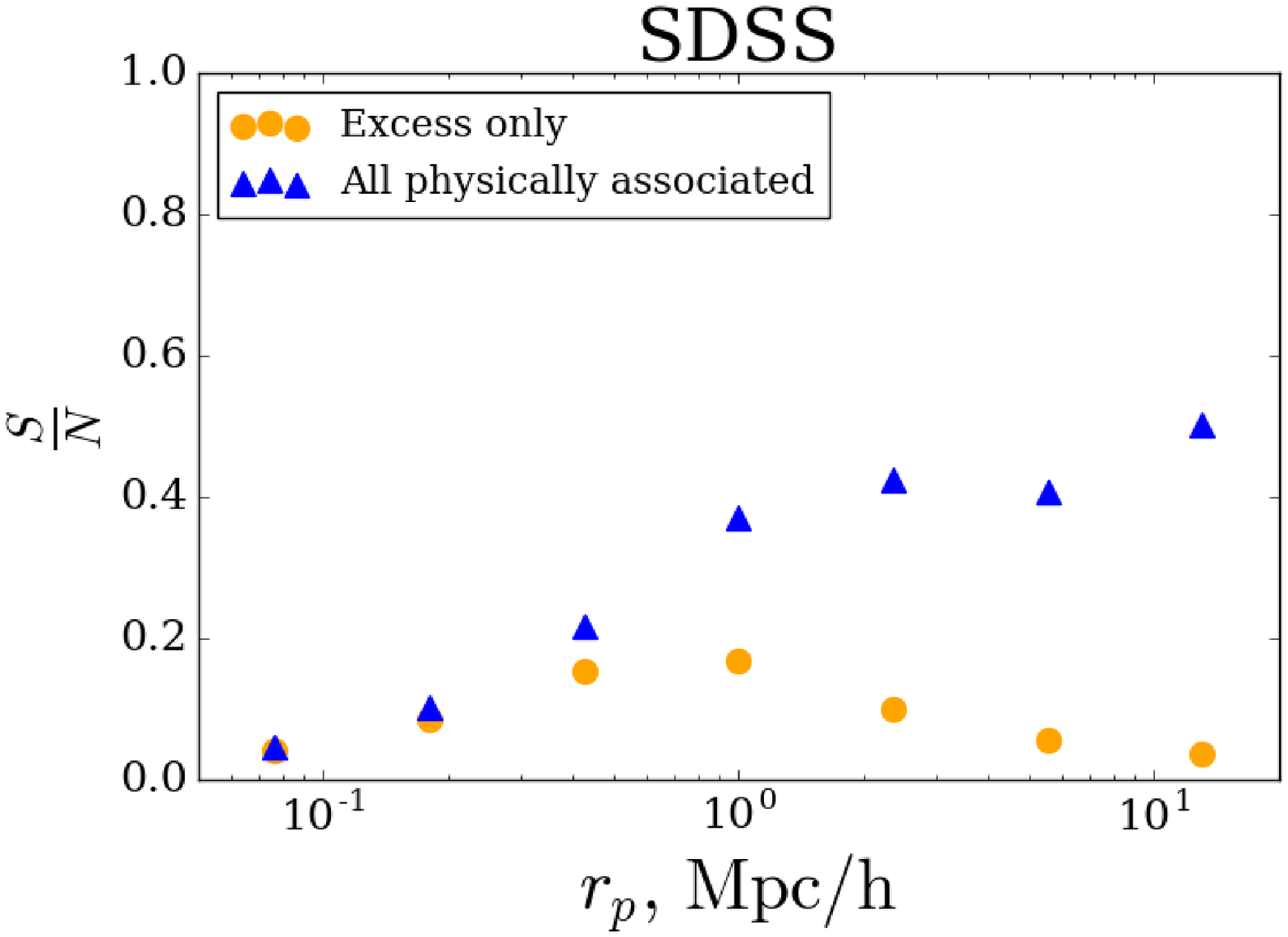}}
\subfigure{\includegraphics[width=0.45\textwidth]{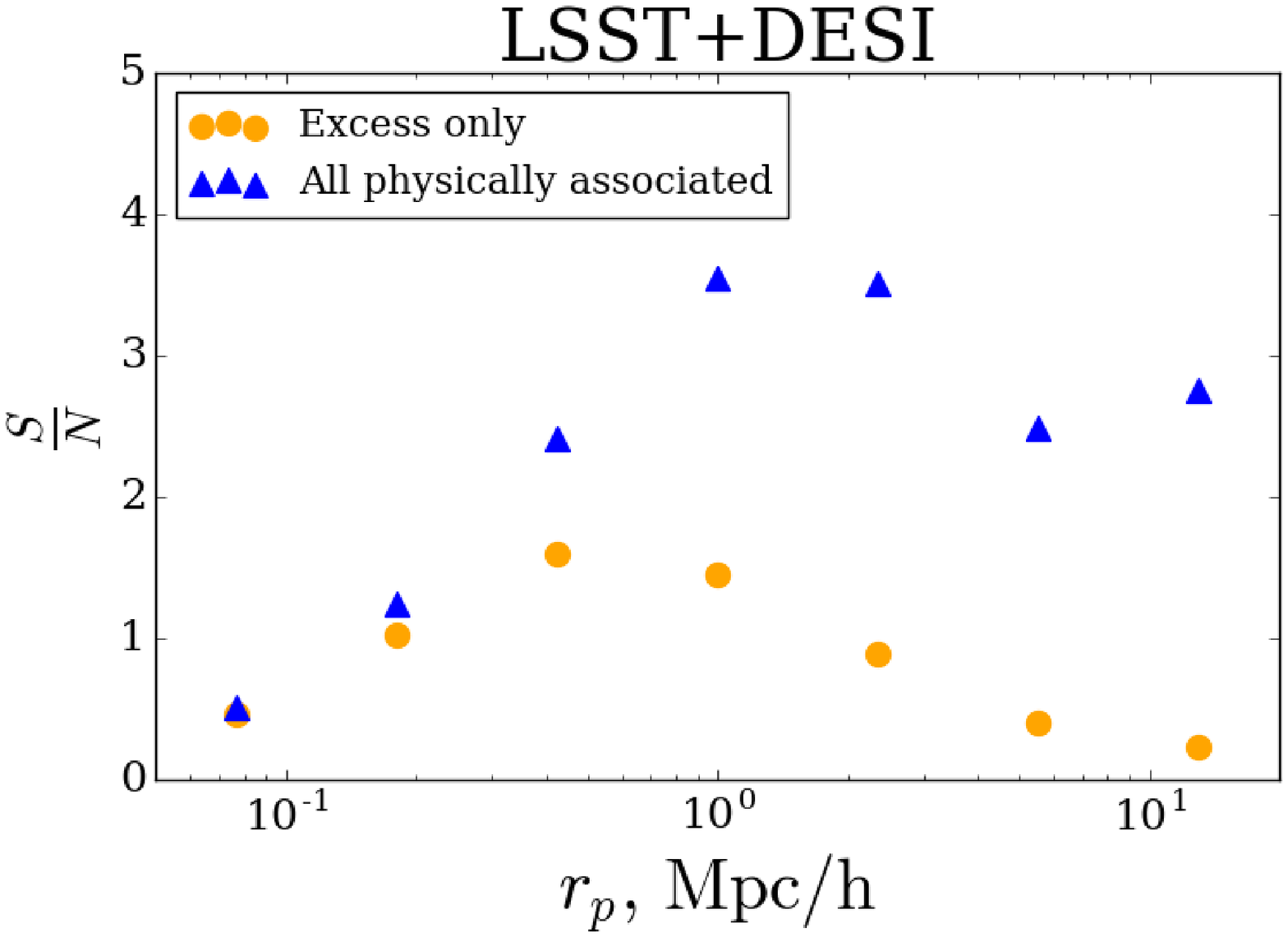}}
\caption{Forecast signal-to-noise on $\bar{\gamma}_{\rm IA}$ using the \protect\cite{Blazek2012} method, including both the original version, which assumes that only excess lens-source pairs are subject to IA, and the modified version developed herein, which assumes that all physically-associated pairs are subject to IA. The latter choice improves the signal-to-noise due to a reduced sensitivity to systematic errors which affect the boost factor. Note that we do not predict a detection of $\bar{\gamma}_{\rm IA}$ in the SDSS case with either version of the \protect\cite{Blazek2012} method, which is in agreement with the measurements of that work.}
\label{fig:Fzerocomparison}
\end{figure*}

Consider, though, for the moment, a scenario in which the systematic error due to source galaxy redshift uncertainty is negligible (an assumption which will be justified below). Uncertainties are then a combination of statistical errors and systematic errors associated with the boost. We can compare the expected constraints on $\bar{\gamma}_{\rm IA}$ from the original version of the method (encapsulated in equation \ref{gammaIA_Blazek}) and this modified version (described in equation \ref{gammaIA_allphys}). The forecast signal-to-noise in each case is displayed in Figure \ref{fig:Fzerocomparison}, for both observational scenarios described in Section \ref{sec:obsscen}. 

We see that the extension to the method of B2012 introduced here improves the signal-to-noise, in particular for the LSST+DESI scenario. This can be understood by noting that the boost factor is subject to a non-negligible systematic error due to effects such as variation of the density of lenses as a result of observational conditions and fluctuations in large-scale clustering (B2012). The boost, representing as it does the weighted ratio of all lens-source pairs to non-excess lens-source pairs, goes to unity on large $r_p$ scales, and so the fractional error associated with the boost increases arbitrarily on these scales, as discussed in B2012. The addition of $F$, which is constant with projected radial separation, ensures that the equivalent term in equation \ref{gammaIA_allphys} never grows arbitrarily large, controlling this error and resulting in the improvement seen in Figure \ref{fig:Fzerocomparison}. When comparing our proposed method to the method of B2012, for the remainder of this work, we use the modification described by equation \ref{gammaIA_allphys}.

\subsection{Redshift-related systematic uncertainties}
\label{subsec:sysresults}
\noindent
Due to the faint limiting magnitude of future lensing surveys such as LSST, potential systematic errors due to inadequately-characterised photo-z uncertainty are a concern \citep{Newman2015}. Quantities affected by this source of error enter our method only via $F$ (see equation \ref{IAsol3}), while in the case of the B2012 method, $c_z^{(i)}$, $\langle \tilde{\Sigma}_c\rangle_{\rm IA}^{(i)}$, and $F_i$ all may be subject to this source of uncertainty. We now compare our method to that of B2012 in terms of robustness to this type of systematic error. 

We first determine the maximum level of redshift-related systematic error which is of practical interest for the measurement of IA contamination to the galaxy-galaxy lensing signal. This is set by the fact that the mitigation of IA is only relevant when the lensing uncertainty itself is not dominated by redshift-related systematics. We therefore work in the regime where the fraction of the total integrated uncertainty on $\Delta \Sigma (r_p)$ which is due to redshift-related systematic error is less than 50\%. We find that in order for this to be true, assuming that the $b$ source sample is representative and typical for a $\Delta \Sigma(r_p)$ measurement, the maximum fractional systematic error on $c_z^{(b)}$ must be less than $\approx 9\times10^{-3}$ in the SDSS case and $\approx 7\times 10^{-4}$ for LSST+DESI. These values in fact also represent upper limits on the redshift-related systematic error of the other six quantities of interest ($c_z^{(a)}$, $F_i$, $\langle \tilde{\Sigma}_c\rangle_{\rm IA}^{(i)}$, $F$), because each of these is sensitive to near-lens sources, for which the photo-z calibration is expected to be best.

We are interested in how the total redshift-related systematic uncertainty on the IA signal compares to its statistical uncertainty, for our method and for the B2012 method. To investigate this, we consider the ratio of $\sigma_{\rm sysz}$ (integrated redshift-related systematic error) to $\sigma_{\rm stat}$ (integrated statistical error); a smaller value of this ratio indicates greater robustness to redshift-related systematic uncertainty, while unity indicates that statistical uncertainty and redshift-related systematic uncertainty are equally important. We examine this ratio as a function of the fractional error on each of $c_z^{(i)}$, $F_i$, $\langle \tilde{\Sigma}_c\rangle_{\rm IA}^{(i)}$, and $F$ in turn (fixing the error on the others to zero), and find a power law relation, with slopes given in Table \ref{table:sysz}. To be explicit, the power law takes the form
\begin{linenomath*}
\begin{equation}
\frac{\sigma_{\rm sysz}}{\sigma_{\rm stat}} = A \left(\frac{\delta x}{x}\right)
\label{powerlawsys}
\end{equation}
\end{linenomath*}
where $x$ is one of $F$, $c_z^{(i)}$, $\langle \tilde{\Sigma}_c\rangle_{\rm IA}^{(i)}$, or $F_i$. We see in Table \ref{table:sysz} that within the B2012 method, the overall importance of redshift-related systematic uncertainty is most sensitive to the level of systematic uncertainty on $c_z^{(a)}$ and $c_z^{(b)}$. This is sensible: the B2012 method incorporates the difference of the two large quantities $c_z^{(a)} \widetilde{\Delta\Sigma}_a$ and $c_z^{(b)} \widetilde{\Delta\Sigma}_b$; uncertainty in either of these terms will result in a relatively large error on their much-smaller difference. 

We use this set of power-law relationships to easily calculate the ratio $\sigma_{\rm sysz} / \sigma_{\rm stat}$ when each of $F$, $c_z^{(i)}$, $\langle \tilde{\Sigma}_c\rangle_{\rm IA}^{(i)}$, or $F_i$ in turn takes its maximum tolerable fractional error (with errors on the remaining quantities fixed at zero). These ratio values are listed in Table \ref{table:sysz_ratios}. We see there that applying the maximum interesting level of redshift-related systematic uncertainty to any individual quantity is insufficient to cause this source of error to dominate over statistical uncertainty.

Even when the redshift-related systematic error on $F$ takes its maximum value of interest, statistical error heavily dominates our method, with $\sigma_{\rm sysz} / \sigma_{\rm stat} \approx 1\%$ for both SDSS and LSST+DESI. For the B2012 method, we examine the worst case scenario in which the redshift-related systematic errors of $c_z^i$, $\langle \tilde{\Sigma}_c\rangle_{\rm IA}^{(i)}$, and $F_i$ all take their maximum values simultaneously, and find the corresponding maximum value of $\sigma_{\rm sysz} / \sigma_{\rm stat}$ to be $29\%$ for SDSS and $5\%$ for LSST+DESI. Although these values are larger than in our method, and thus we see that the B2012 method is less robust to redshift-related systematic errors in the worst-case scenario, statistical error remains dominant.

\begin{center}
\begin{table}
\begin{center}
\begin{tabular}{| c | c | c |c|}
\hline
&\begin{tabular}[c]{@{}c@{}}Slope $A$,\\SDSS\end{tabular}&\begin{tabular}[c]{@{}c@{}}Slope $A$,\\LSST+DESI\end{tabular}& \,Eqn \, \\
\hline
\vspace{1mm}
\,\, $c_z^{(a)}$ \,\,  & \,\, 23 \,\, & \,\, 48 \,\, & \,\, \ref{photozbias} \,\, \\
\vspace{1mm}
\,\, $c_z^{(b)}$\,\,  & \,\, 22 \,\, & \,\, 41 \,\, & \,\, \ref{photozbias} \,\, \\
\vspace{1mm}
\,\, $F_a$\,\,  & \,\, 0.72 \,\, &\,\,  4.3 \,\, & \,\, \ref{F_def} \,\,\\
\vspace{1mm}
\,\, $F_b$ \,\, &\,\,  0.063 \,\, & \,\,  0.012\,\, & \,\, \ref{F_def} \,\, \\
\vspace{1mm}
\,\, $\langle \tilde{\Sigma}_c\rangle_{\rm IA}^{(a)}$ \,\, & \,\, 0.95 \,\, & \,\, 6.5 \,\, & \,\, \ref{SigIA} \,\, \\
\vspace{1mm}
\,\, $\langle \tilde{\Sigma}_c\rangle_{\rm IA}^{(b)}$\,\,  &\,\, 0.081\,\,  & \,\, 0.018\,\, & \,\, \ref{SigIA} \,\, \\
\vspace{1mm}
\,\, $F$ \,\, & \,\, 1.2\,\,  & \,\, 8.1\,\, & \,\, \ref{F_def} \,\, \\
\hline
\end{tabular}
\end{center}
\vspace{-0.2em}
\caption{Slope of the linear relationship between the ratio of total integrated (redshift-related) systematic to statistical error on the IA signal and the fractional level of systematic error on the quantity in the leftmost column. A larger value indicates that the importance of systematic error in the total error budget depends more sensitively on the systematic error on the given quantity.}
\vspace{-1.2em}
\label{table:sysz}
\end{table}
\end{center}

\begin{center}
\begin{table}
\begin{center}
\begin{tabular}{| c | c | c |c|}
\hline
&\begin{tabular}[c]{@{}c@{}}Max $\frac{\sigma_{\rm sysz}}{\sigma_{\rm stat}}$\\SDSS\end{tabular}&\begin{tabular}[c]{@{}c@{}}Max $\frac{\sigma_{\rm sysz}}{\sigma_{\rm stat}}$\\LSST+DESI\end{tabular}& \,Eqn \, \\
\hline
\vspace{1mm}
\,\, $c_z^{(a)}$ \,\,\,\,  & \,\, 0.21 \,\, & \,\, 0.035 \,\, & \,\, \ref{photozbias} \,\, \\
\vspace{1mm}
\,\, $c_z^{(b)}$\,\,  & \,\,  0.20 \,\, & \,\, 0.030 \,\, & \,\, \ref{photozbias} \,\, \\
\vspace{1mm}
\,\, $F_a$\,\,  & \,\, $6.6 \times 10^{-3}$ \,\, &\,\, $3.2\times 10^{-3}$  \,\, & \,\, \ref{F_def} \,\,\\
\vspace{1mm}
\,\, $F_b$ \,\, &\,\, $5.8 \times 10^{-4}$ \,\, & \,\, $9.1\times 10^{-6}$ \,\, & \,\, \ref{F_def} \,\, \\
\vspace{1mm}
\,\, $\langle \tilde{\Sigma}_c\rangle_{\rm IA}^{(a)}$ \,\, & \,\, $8.8\times 10^{-3}$ \,\, & \,\, $4.8 \times 10^{-3}$ \,\, & \,\, \ref{SigIA} \,\, \\
\vspace{1mm}
\,\, $\langle \tilde{\Sigma}_c\rangle_{\rm IA}^{(b)}$\,\,  &\,\, $7.5 \times 10^{-4}$ \,\,  & \,\, $1.3\times 10^{-5}$ \,\, & \,\, \ref{SigIA} \,\, \\
\vspace{1mm}
\,\, $F$ \,\, & \,\, 0.011 \,\,  & \,\, $5.9\times10^{-3}$ \,\, & \,\, \ref{F_def} \,\, \\ 
\hline
\end{tabular}
\end{center}
\vspace{-0.2em}
\caption{Maximum expected value of the ratio of total integrated (redshift-related) systematic to statistical error on the IA signal, given that the quantity in the leftmost column carries the maximum tolerable redshift-related systematic uncertainty and each other quantity carries none. A value of less than unity indicates that statistical error is dominant.}
\vspace{-1.2em}
\label{table:sysz_ratios}
\end{table}
\end{center}

\subsection{Statistical uncertainty}
\label{subsec:statresults}
\noindent
As we have shown in Section \ref{subsec:sysresults}, the uncertainty on $\bar{\gamma}_{\rm IA}$ in our proposed method (and in the method of B2012) cannot realistically be dominated by redshift-related systematic errors. With this in mind, we now examine the relative performance of these methods when statistical errors dominate.

We compute the signal-to-noise of the intrinsic alignment measurement from each method, assuming only statistical uncertainty and a small contribution due to systematic error on the boost factor, $\sigma_B = 0.03$ (B2012) (see Appendix \ref{app:cov} for how the required covariance matrices are calculated in each scenario). Note that for both methods, we assume that statistical uncertainty on the boost is negligible; we have tested this assumption and found that including this source of error yields less than a $1\%$ change in the total uncertainty. For our method, we compute the signal-to-noise as a function of $a$, the ratio of intrinsic alignment amplitudes between the two shape measurement methods, and of the correlation between the shape-noise of the shear estimates, which we call $\rho$. As stated above, $a\approx 0.7-0.8$ for the shear estimation methods found in \cite{Singh2016} (isophotal shapes, de Vaucouleurs shapes, and re-Gaussianization). The same work finds $\rho \approx 0.7$ for pairs of these methods, as quoted in \cite{Chisari2016}. However, as previously stated we imagine our proposed method to be most useful in the context of a modified Bayesian Fourier Domain method with custom radial weighting, which could in principle allow for the construction of shear estimates with a wide variety of $a$ and $\rho$ values.

\begin{figure*}
\centering
\subfigure{\includegraphics[width=0.45\textwidth]{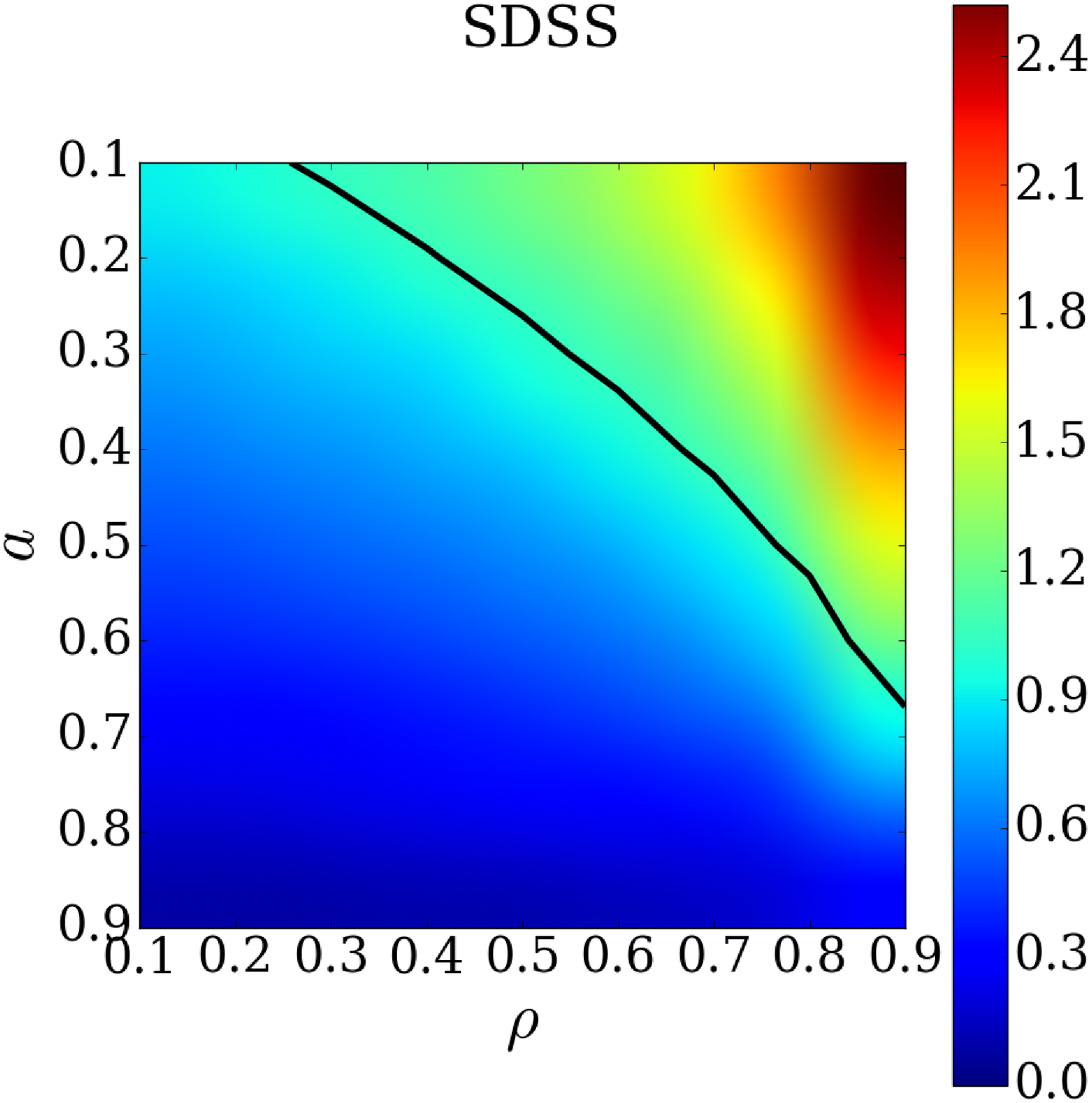}}
\subfigure{\includegraphics[width=0.45\textwidth]{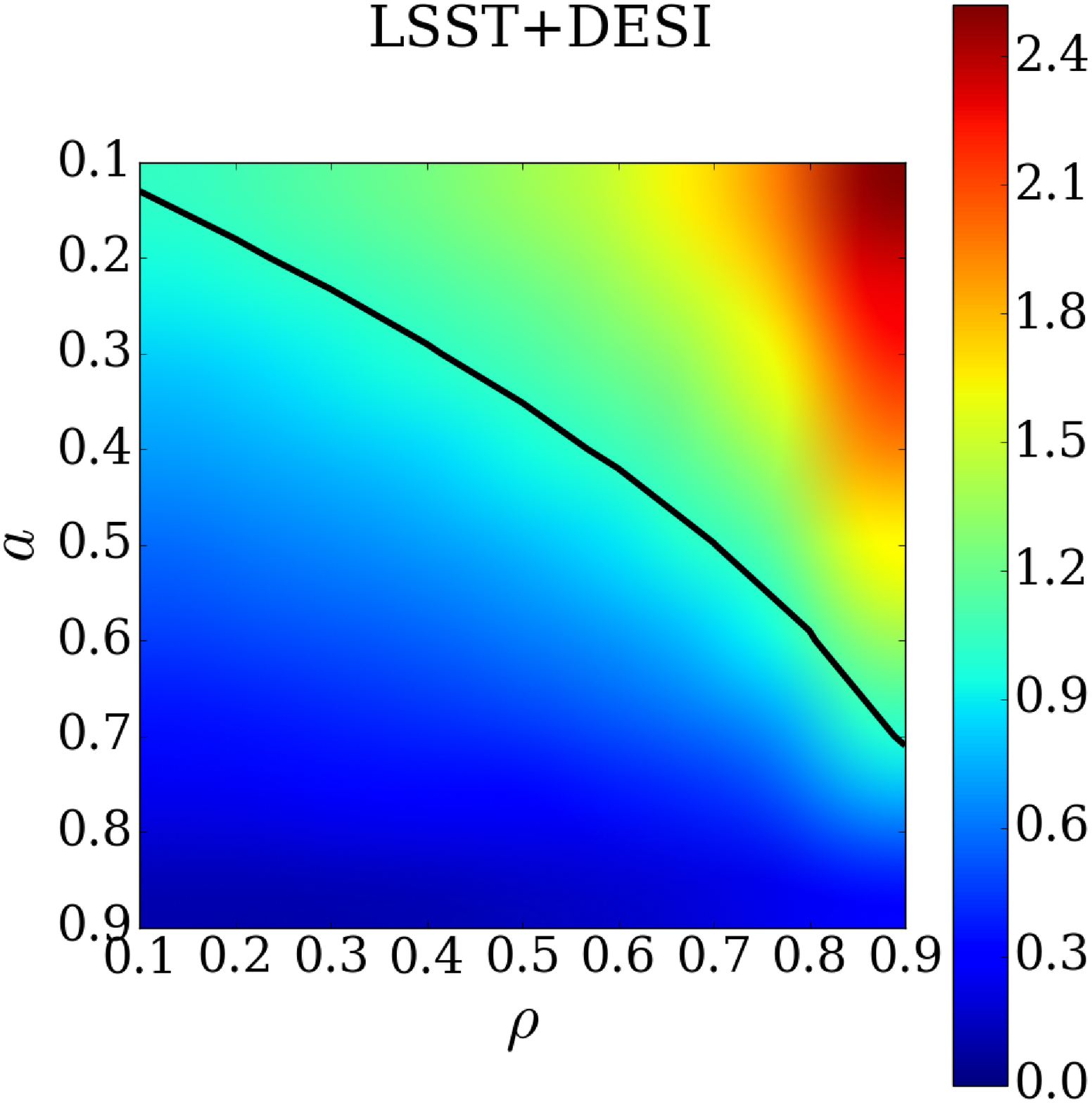}}
\caption{Integrated signal-to-noise in our proposed method, divided by integrated signal-to-noise in the method of \protect\cite{Blazek2012}, under the assumption that redshift-related systematic uncertainties are subdominant. $a$ represents the difference in measured IA amplitude between shear estimates (where smaller $a$ is a greater difference), and $\rho$ is the correlation between the shape noise of each shear estimate. Signal-to-noise is reported for signal of $\bar{\gamma}_{\rm IA}$ or $(1-a)\bar{\gamma}_{\rm IA}$ as appropriate for the method, and is integrated over scales $r_p=0.05-20$ Mpc/h. A value of unity or greater indicates that the proposed method out-performs the existing method; the black curve indicates the contour where the ratio is equal to one, with values greater than unity to the upper right corner in both cases.}
\label{fig:StoNstat}
\end{figure*}

In Figure \ref{fig:StoNstat}, we plot the forecast ratio of the integrated signal-to-noise of our method to the signal-to-noise of the B2012 method as a function of $a$ and $\rho$, for both SDSS and LSST+DESI. A value greater than unity indicates that for the given $(a,\,\rho)$ pair, the proposed method out-performs the existing one. Our method is shown to perform best for lower values of $a$ and for higher values of $\rho$. Lower values of $a$ correspond to pairs of shear estimates which produce more divergent intrinsic alignment amplitudes, and therefore which increase the signal $(1-a)\bar{\gamma}_{\rm IA}$. Higher values of $\rho$ indicate increased correlation between the shape-noise terms of the two shear estimates, which results in a reduced level of noise on the difference of tangential shear estimators. 

The segment of the ($a$, $\rho$) plane for which our method performs better than that of B2012 is larger in the LSST+DESI scenario, the reason for which is not entirely obvious and warrants mention. The reduction in photo-z uncertainty in going from SDSS to LSST+DESI means that for a given source sample, less spurious sources are inadvertently included. In our method, this means that a yet-higher proportion of pairs are subject to intrinsic alignments. Similarly, the effective fraction of pairs subject to IA in source sample $b$ of the B2012 method drops. Less intuitively, the effective proportion of pairs subject to IA in the $a$ sample of B2012 also drops, because this method down-weights pairs which are close in photometric redshift, of which there are more in the $a$ sample when photo-z uncertainty is reduced. Using redshift-independent weights in a B2012-like method does not entirely remove this effect, as it is dominated by the decrease in IA-subject lens-source pairs in the $b$ sample. The choice of $\Delta z$ for the B2012 method also leads to some perceived increase in our method's performance in the LSST+DESI scenario as compared to SDSS, which is an artifact of the fact that we adopt, in the SDSS case, a value of $\Delta z$ that jointly optimises for lensing and IA measurement. For LSST+DESI, we select $\Delta z=0.1$ to optimise the signal-to-noise of $\bar{\gamma}_{\rm IA}$, while in the SDSS case, we have set $\Delta z = 0.17$ to follow the methodology of B2012. The redshift extent of source sample $a$, the noise of which dominates the B2012 method, is therefore decreased in moving from the SDSS to the LSST+DESI scenario, while that in our method remains roughly constant.

\begin{figure*}
\centering
\includegraphics[width=0.45\textwidth]{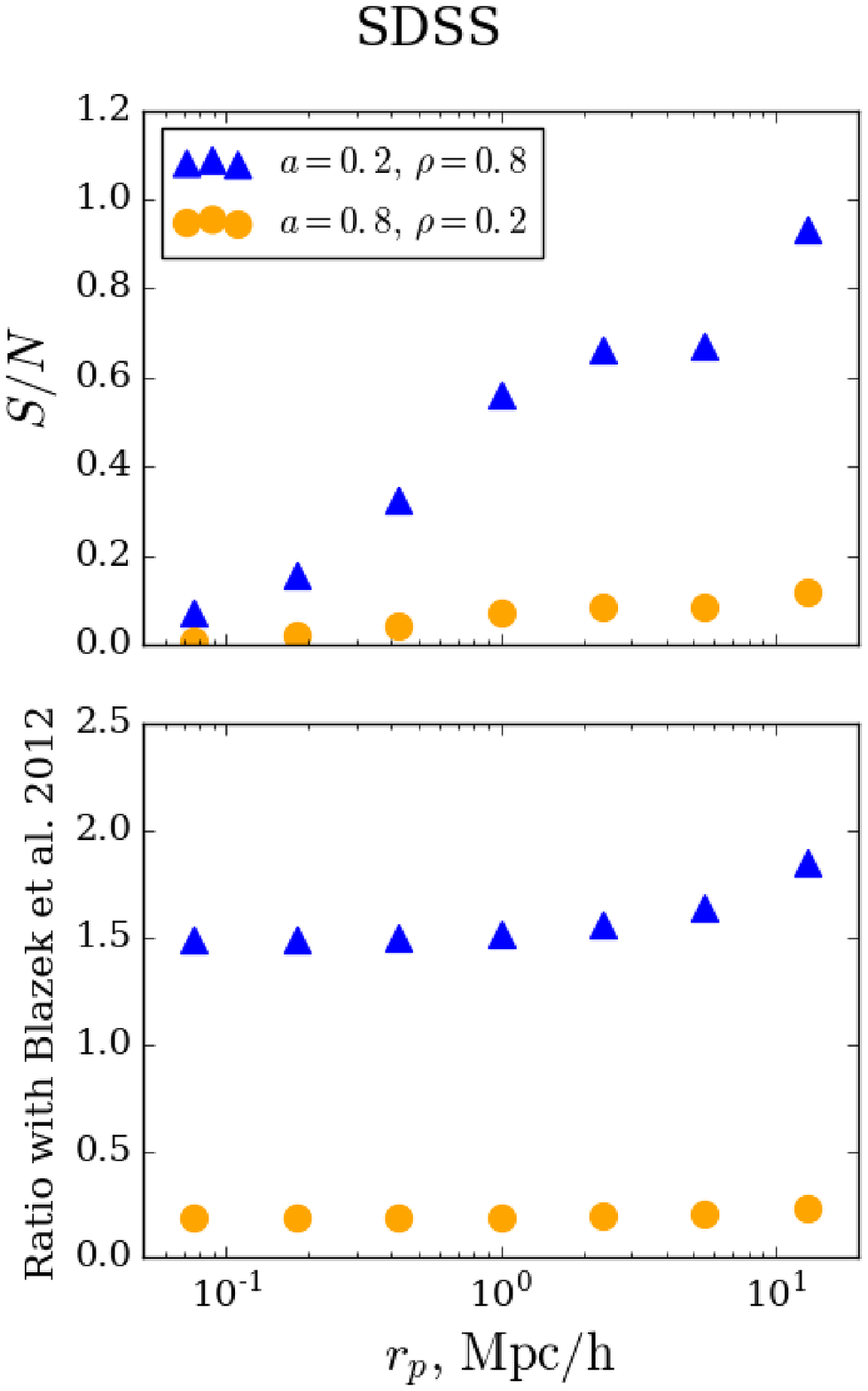}
\includegraphics[width=0.45\textwidth]{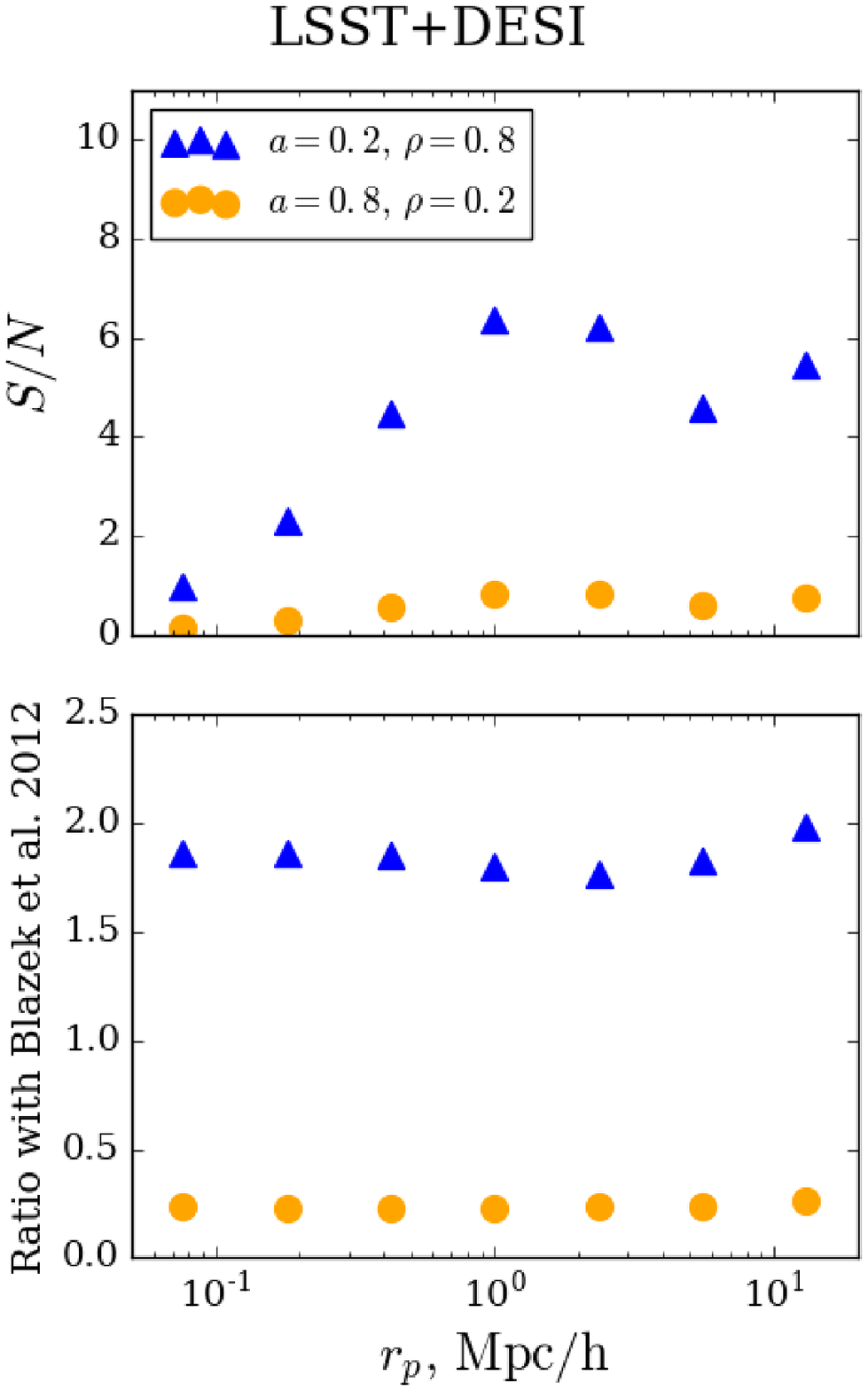}
\caption{Top panel: signal-to-noise as a function of projected radius in our proposed method, under the assumption that redshift-related systematic uncertainties are subdominant, for two sample pairs of $a$ and $\rho$. The characteristic scale dependence of the signal-to-noise within a given survey does not change significantly with the values of $a$ and $\rho$. Lower panel: the same, divided by signal-to-noise in the \protect\cite{Blazek2012} method. Signal-to-noise is reported for signal of $\bar{\gamma}_{\rm IA}$ or $(1-a)\bar{\gamma}_{\rm IA}$ as appropriate for the method.}
\label{fig:StoNstat_1d}
\end{figure*}

In Figure \ref{fig:StoNstat_1d}, we show the signal-to-noise in our method as a function of $r_p$, as well as the ratio of signal-to-noise in our method to that in the B2012 method, for two sets of $(a$, $\rho)$ values: $(a=0.8, \, \rho=0.2)$ and $(a=0.2, \, \rho=0.8)$. The former is an example of the region of parameter space in which the B2012 method is superior for both observational scenarios, while the latter is a case for which our proposed method performs significantly better. For $(a=0.2, \, \rho=0.8)$ and in the LSST+DESI scenario, we forecast an integrated signal-to-noise for our method of 11 (integrated over $r_p=0.05-20$ Mpc/h). This represents a significant improvement over the existing method, for which we predict a signal-to-noise of 6. Thus, given two shear estimates with these characteristic values of $a$ and $\rho$, our method has the potential to allow useful inferences about the signal for modelling purposes. Additionally, we have computed these forecast signal-to-noise values assuming a source sample consisting of all LSST source galaxies down to $r_{\rm lim}=25.3$; with restriction to a subset with a less-faint limiting magnitude and hence higher IA contamination, yet a higher signal-to-noise could presumably be achieved. Finally, we note from Figure \ref{fig:StoNstat_1d} that the relative signal-to-noise of our method is higher at larger $r_p$. This scale dependence is characteristic for all $(a,\,\rho)$ pairs, and suggests that our method may be of the most use in making measurements which inform the modelling of the transitional regime in which one-halo and two-halo contributions are both relevant.

\section{Discussion and Conclusions}
\label{sec:conclusions}
In this work, we have proposed a new method of measuring the scale dependence of the intrinsic alignment contribution to the galaxy-galaxy lensing signal. Our method takes advantage of the result of \cite{Singh2016}, who found that shear estimates with sensitivity to different radial parts of the light profile of galaxies produce intrinsic alignment signals with amplitudes which are offset by a constant multiplicative factor.

Comparing our method to the existing method of B2012, we find that for considerable portions of the ($a$, $\rho$) parameter space, the proposed method is forecast to out-perform the existing method when redshift-related systematic errors are controlled. The improvement of signal-to-noise in our method as compared to the existing method is greater for the LSST+DESI observational scenario than for the SDSS case, primarily due to the reduction in photometric redshift uncertainties. This suggests that this method may be especially fruitful in mitigating intrinsic alignment effects in future surveys. This promising situation merits a more detailed investigation, using appropriate simulated datasets.

Our method is found to be more robust to systematic uncertainties related to difficulties in characterising photo-z errors than is the existing method to which we compare. However, for realistic levels of this redshift-related systematic error (for which this source of uncertainty does not dominate measurements of the lensing signal itself), the IA measurement of both our method and the method of B2012 are necessarily dominated by statistical uncertainty.

The signal-to-noise of the IA measurement forecast by our method in the regime in which redshift-related systematic errors are controlled depends on $a$ and $\rho$. Given that it is possible to predict for which segments of this parameter space our method's signal-to-noise is best (see Figure \ref{fig:StoNstat}), the possibility arises of constructing pairs of shear estimates for which the values of $a$ and $\rho$ optimise this signal-to-noise. This could in principle be accomplished through minor modifications to the Bayesian Fourier Domain (BFD) method \citep{Bernstein2014, Bernstein2016}, into which multiple radial weighting functions could potentially be incorporated to create shear estimation methods with custom values of $a$ and $\rho$. Currently, the BFD method is under development, with further work required to create mature estimators for two-point functions within this framework as well as to deal with other effects such as, for example, blending. However, it is a promising shear-estimation framework and one which could be combined with the method proposed in this work to great effect. Another option of potential interest would be to use a BFD-type shear estimate and a metacalibration shear estimate \citep{Huff2017, Sheldon2017}, provided the two could be constructed in such a way that their radial weightings were different. However, in this latter scenario, it would be necessary to somehow account for any differing selection effects, which if unmitigated could negate the essential cancellation of the lensing signal.

One particularly interesting scenario in which the proposed method could potentially be employed is the case of a joint analysis of weak lensing data from a ground-based survey such as LSST and a space-based survey such as WFIRST or Euclid. The atmospheric point spread function associated with ground-based surveys tends to prevent the use of the smallest radii of the light profile in shear estimation, whereas for space-based imaging this issue is avoided. Given an overlap in sky area and a similar limiting magnitude, one could therefore imagine measuring $\tilde{\gamma}_t$ from LSST and $\tilde{\gamma}_t^\prime$ from WFIRST or Euclid. The object detection and deblending could be carried out in the space-based survey, avoiding selection-related biases. In this application of the proposed method, the two shear estimation methods could have maximally different radial sensitivity, due to the fact that in the case of measuring $\tilde{\gamma}_t^\prime$ from the space-based survey, the chosen shear estimation method could be sensitive at smaller radii than would be possible in ground-based imaging.

In order to make a fair comparison between our proposed method and the method of B2012, we introduced a modification to the latter such that we assume that all physically-associated lens-source pairs are subject to IA, as opposed to only excess pairs. We find that this modification improves the signal-to-noise of this method. Although this modification may appear at first glance to introduce new sources of redshift-related systematic uncertainties, we have also shown, as discussed, that these types of systematic uncertainties must be subdominant. 

Our proposed method is forecast to improve upon the signal-to-noise of the existing method to which we compare for the LSST+DESI observational scenario, presenting a promising situation which invites further investigation. For example, a straightforward practical exploration of this method could be made by considering isophote estimators at different galactic radii. It would be interesting to investigate initially the relatively simple case of an analytic galaxy profile parameterisation, in order to gain insight as to the expected range of $a$ values. Additionally, as our method is intended to primarily constrain the scale dependence of the IA signal while being largely insensitive to amplitude, it would benefit from combination with an existing method to enable a full IA signal measurement. One obvious avenue would be to extend our proposed method to incorporate elements of the method of B2012, potentially providing an even more powerful probe of IA. Furthermore, in this work, we have not derived the redshift-dependence of weights which would optimise the signal-to-noise of our measurement, rather assuming a typical redshift-independent choice. With such an optimal weighting, we anticipate that the signal-to-noise of the method we present would be yet higher.

Several upcoming cosmological surveys, including LSST, as discussed in this work, but also Euclid, WFIRST, and others, will soon engender a radical reduction in statistical uncertainties of lensing measurements. Understanding the intrinsic alignment contribution to lensing signals and mitigating its effect will thus become critically important, as percent-level effects such as this will begin to have significant relevance. The new method of measuring the scale dependence of the intrinsic alignment signal which we have presented here may be of great assistance, as it has the capacity to perform significantly better than the existing method to which we compare for a next-generation galaxy-galaxy lensing measurement.

\section*{Acknowledgements}
This paper has undergone internal review in the LSST Dark Energy Science Collaboration. The authors thank Joe Zuntz, Fran\c{c}ois Lanusse, Sukhdeep Singh, Melanie Simet, Michael Troxel, Benjamin Joachimi, Gary Bernstein, Simon Samuroff, and Jonathan Blazek for helpful discussions, as well as the LSST Dark Energy Science Collaboration internal review panel of Tim Eifler, Patricia Larsen, and Michael Schneider, for valuable feedback and suggestions. CDL is supported by a McWilliams Postdoctoral Fellowship. RM is supported by the Department of Energy Cosmic Frontier program. The DESC acknowledges ongoing support from the Institut National de Physique Nucl\'eaire et de Physique des Particules in France; the Science \& Technology Facilities Council in the United Kingdom; and the Department of Energy, the National Science Foundation, and the LSST Corporation in the United States. DESC uses resources of the IN2P3 Computing Center (CC-IN2P3--Lyon/Villeurbanne - France) funded by the Centre National de la Recherche Scientifique; the National Energy Research Scientific Computing Center, a DOE Office of Science User Facility supported by the Office of Science of the U.S.\ Department of Energy under Contract No.\ DE-AC02-05CH11231; STFC DiRAC HPC Facilities, funded by UK BIS National E-infrastructure capital grants; and the UK particle physics grid, supported by the GridPP Collaboration. This work was performed in part under DOE Contract DE-AC02-76SF00515. The Python libraries SciPy (\citealt{scipy}) and NumPy (\citealt{numpy}) were used in this work.

Author contributions: CDL performed all analysis (including derivations, writing code, and validation of code), and wrote the text of the paper. RM conceptualised the project, provided advice and suggestions, and edited the text of the paper.





\appendix

\section{Covariance matrices of $\tilde{\gamma}_{\lowercase{t}}$ and $\widetilde{\Delta \Sigma}$}
\label{app:cov}
In this appendix we provide the required theoretical expressions to compute the statistical covariance matrices used to produce the results of Section \ref{subsec:statresults}.

\subsection{$\widetilde{\Delta \Sigma}$}
The covariance of $\widetilde{\Delta \Sigma}$ in projected radial bin $r_p^i$ and $r_p^j$ is given by (see, for example, \citep{Singh2016b, Jeong2009}):
\begin{linenomath*}
\begin{align}
{\rm Cov}&\left[\widetilde{\Delta \Sigma}\left(r_p^i\right),\widetilde{\Delta \Sigma}\left(r_p^j\right) \right] = \frac{1}{4\pi f_{ \rm sky}} \left(\overline{\Sigma_c^{-2}}\right)^{-1} \nonumber \\ &\times \frac{2}{\left(\left(r_p^{i,h}\right)^2-\left(r_p^{i,l}\right)^2 \right)} \frac{2}{\left(\left(r_p^{j,h}\right)^2-\left(r_p^{j,l}\right)^2\right) }   \nonumber \\ & \times \int_{r_p^{i,l}}^{r_p^{i,h}} dr_p \, r_p \int_{r_p^{j,l}}^{r_p^{j,h}} dr_p^\prime \, r_p^\prime \int \frac{l dl}{2\pi} J_2\left(l\frac{r_p}{\chi\left(z_{{l}}^{\rm eff}\right)}\right)\nonumber \\ &\times J_2\left(l\frac{r_p^\prime}{\chi\left(z_{{l}}^{\rm eff}\right)}\right)\Bigg[\left(C_{g\kappa}^{l}\right)^2 + \left(C_{gg}^l+\frac{1}{n_{l}}\right)\Bigg( C_{\kappa\kappa}^l + \frac{\sigma_{\gamma}^2}{n_{eff}} \Bigg)\Bigg],
\label{DeltaSigmaCov}
\end{align}
\end{linenomath*}
where $r_p^{i,h}$ and $r_p^{i,l}$ represent the high and low edges of bin $r_p^i$ respectively (and similarly for bin $r_p^j$), $n_{l}$ is the surface density of lens galaxies, $n_{eff}$ is the effective surface density of source galaxies (both in galaxies per steradian), and $\overline{\Sigma_c^{-2}}$ is given by:
\begin{linenomath*}
\begin{equation}
\overline{\Sigma_c^{-2}} = \int dz_{l} \frac{dN}{dz_{l}} \int dz_{ph} \frac{dN}{dz_{ph}} \Sigma_c^{-2}(z_{l}, z_{ph}). 
\label{wbar}
\end{equation}
\end{linenomath*}
The angular power spectra in equation \ref{DeltaSigmaCov} should be computed for the source and lens samples of interest, incorporating in the case of sources cuts on photometric redshifts. Note that we integrate over the full lens redshift distribution everywhere except in the argument of the Bessel functions. In this case, we use the comoving distance corresponding to the effective lens population redshift. We do not expect this to affect results significantly. 

We use equation \ref{DeltaSigmaCov} in calculating the statistical uncertainty and covariance between $r_p$ bins of $\bar{\gamma}_{\rm IA}(r_p)$ for the existing method of B2012. A covariance matrix of this form is computed independently for each source sample, $a$ and $b$, and for each of the two observational scenarios. Because source samples $a$ and $b$ do not have significant overlap, the covariance matrices for $\widetilde{\Delta \Sigma}_a(r_p)$ and $\widetilde{\Delta \Sigma}_b(r_p)$ are considered independent and are combined under this assumption. Practically, when performing this computation, we take advantage of the orthogonality of the Bessel functions to separate out of the constant shape-noise term.

Note that we include one-halo terms in the calculation of the power spectra in equation \ref{DeltaSigmaCov}; this is necessary particularly in the case of the LSST+DESI observational scenario, as for this case shape noise is sufficiently low that cosmic variance dominates on some scales in the one-halo regime.

\subsection{$\tilde{\gamma}_t$}

The covariance matrix for $\tilde{\gamma}_t$, which is required for the computation of the statistical contribution to the covariance matrix for the new method proposed in this work, is given by an expression very similar to equation \ref{DeltaSigmaCov}, different only in factors of the critical surface mass density:
\begin{linenomath*}
\begin{align}
{\rm Cov}&\left[\tilde{\gamma}_t\left(r_p^i\right),\tilde{\gamma}_t \left(r_p^j\right) \right] = \frac{1}{4\pi f_{ \rm sky}}\nonumber \\ &\times \frac{2}{\left(\left(r_p^{i,h}\right)^2-\left(r_p^{i,l}\right)^2 \right)} \frac{2}{\left(\left(r_p^{j,h}\right)^2-\left(r_p^{j,l}\right)^2\right) }   \nonumber \\ & \times \int_{r_p^{i,l}}^{r_p^{i,h}} dr_p \, r_p \int_{r_p^{j,l}}^{r_p^{j,h}} dr_p^\prime \, r_p^\prime \int \frac{l dl}{2\pi} J_2\left(l\frac{r_p}{\chi\left(z_{{l}}^{\rm eff}\right)}\right)\nonumber \\ &\times J_2\left(l\frac{r_p^\prime}{\chi\left(z_{{l}}^{\rm eff}\right)}\right)\Bigg[\left(C_{g\kappa}^{l}\right)^2 + \left(C_{gg}^l+\frac{1}{n_{l}}\right)\Bigg( C_{\kappa\kappa}^l + \frac{\sigma_{\gamma}^2}{n_{eff}} \Bigg)\Bigg].
\label{GammaCov}
\end{align}
\end{linenomath*}

In this case, we ultimately require the covariance matrix for $\tilde{\gamma}_t - \tilde{\gamma}_t^\prime$ in projected radial bins. Because $\tilde{\gamma}_t$ and $\tilde{\gamma}_t^\prime$ are measured from the same set of lens-source pairs, terms in the above expression which depend only on the galaxy-galaxy lensing signal (or which depend only on the signal and the shot noise of the lenses, i.e. the term proportional to $C^l_{\kappa\kappa} / n_{l}$), are fully correlated. These terms therefore subtract off entirely when computing the covariance of $\tilde{\gamma}_t - \tilde{\gamma}_t^\prime$. Terms which are related to the shape noise (i.e. the terms proportional to $C^l_{gg} \sigma_\gamma^2/ n_{eff}$ and $\sigma_{\gamma}^2 / (n_{eff} n_{l})$), are partially correlated, where the degree of correlation of these terms depends on the shear estimation methods in question, and is parameterised by $\rho$, as discussed in Section \ref{subsec:statresults}. The resulting covariance matrix is given by:
\begin{linenomath*}
\begin{align}
{\rm Cov}&\left[\tilde{\gamma}_t\left(r_p^i\right) - \tilde{\gamma}_t^\prime\left(r_p^i\right),\tilde{\gamma}_t\left(r_p^j\right) - \tilde{\gamma}_t^\prime\left(r_p^j\right) \right]= \frac{1}{4\pi f_{ \rm sky}}\nonumber \\ &\times \frac{2}{\left(\left(r_p^{i,h}\right)^2-\left(r_p^{i,l}\right)^2 \right)} \frac{2}{\left(\left(r_p^{j,h}\right)^2-\left(r_p^{j,l}\right)^2\right) }   \nonumber \\ & \times \int_{r_p^{i,l}}^{r_p^{i,h}} dr_p \, r_p \int_{r_p^{j,l}}^{r_p^{j,h}} dr_p^\prime \, r_p^\prime \int \frac{l dl}{2\pi} J_2\left(l\frac{r_p}{\chi\left(z_{{l}}^{\rm eff}\right)}\right)\nonumber \\ &\times J_2\left(l\frac{r_p^\prime}{\chi\left(z_{{l}}^{\rm eff}\right)}\right) \frac{C_{gg}^l}{n_{eff}} \left(\sigma_\gamma^2 + \left(\sigma_\gamma^\prime\right)^2 - 2 \rho \sigma_\gamma \sigma_\gamma^\prime \right) \nonumber \\ &+ \delta_{ij} \frac{\chi\left(z_{{l}}^{\rm eff}\right)^2 \left(\sigma_\gamma^2 +\left(\sigma_\gamma^\prime\right)^2 - 2 \rho \sigma_\gamma \sigma_\gamma^\prime \right)}{4 \pi^2 f_{\rm sky} \left[\left(r_p^{i,h}\right)^2-\left(r_p^{i,l}\right)^2\right]n_{eff} n_{l}}
\label{GammaCov_diff}
\end{align}
\end{linenomath*}
where we have this time explicitly separated off the shape-noise term by taking advantage the orthogonality of the Bessel functions. In this work we have had $\sigma_\gamma = \sigma_\gamma^\prime$ in both scenarios considered, but this need not be the case and is left general in equation \ref{GammaCov_diff}.


\bsp	
\label{lastpage}

\begin{thebibliography}{}
\makeatletter
\relax
\def\mn@urlcharsother{\let\do\@makeother \do\$\do\&\do\#\do\^\do\_\do\%\do\~}
\def\mn@doi{\begingroup\mn@urlcharsother \@ifnextchar [ {\mn@doi@}
  {\mn@doi@[]}}
\def\mn@doi@[#1]#2{\def\@tempa{#1}\ifx\@tempa\@empty \href
  {http://dx.doi.org/#2} {doi:#2}\else \href {http://dx.doi.org/#2} {#1}\fi
  \endgroup}
\def\mn@eprint#1#2{\mn@eprint@#1:#2::\@nil}
\def\mn@eprint@arXiv#1{\href {http://arxiv.org/abs/#1} {{\tt arXiv:#1}}}
\def\mn@eprint@dblp#1{\href {http://dblp.uni-trier.de/rec/bibtex/#1.xml}
  {dblp:#1}}
\def\mn@eprint@#1:#2:#3:#4\@nil{\def\@tempa {#1}\def\@tempb {#2}\def\@tempc
  {#3}\ifx \@tempc \@empty \let \@tempc \@tempb \let \@tempb \@tempa \fi \ifx
  \@tempb \@empty \def\@tempb {arXiv}\fi \@ifundefined
  {mn@eprint@\@tempb}{\@tempb:\@tempc}{\expandafter \expandafter \csname
  mn@eprint@\@tempb\endcsname \expandafter{\@tempc}}}

\bibitem[\protect\citeauthoryear{Ade et~al.,}{Ade et~al.}{2016}]{Planck2015}
Ade P.~A.,  et~al., 2016, Astronomy \& Astrophysics, 594, A13

\bibitem[\protect\citeauthoryear{{Bernstein} \& {Armstrong}}{{Bernstein} \&
  {Armstrong}}{2014}]{Bernstein2014}
{Bernstein} G.~M.,  {Armstrong} R.,  2014, \mn@doi [\mnras]
  {10.1093/mnras/stt2326}, \href
  {http://adsabs.harvard.edu/abs/2014MNRAS.438.1880B} {438, 1880}

\bibitem[\protect\citeauthoryear{{Bernstein} \& {Jarvis}}{{Bernstein} \&
  {Jarvis}}{2002}]{Bernstein2002}
{Bernstein} G.~M.,  {Jarvis} M.,  2002, \mn@doi [\aj] {10.1086/338085}, \href
  {http://adsabs.harvard.edu/abs/2002AJ....123..583B} {123, 583}

\bibitem[\protect\citeauthoryear{{Bernstein}, {Armstrong}, {Krawiec}  \&
  {March}}{{Bernstein} et~al.}{2016}]{Bernstein2016}
{Bernstein} G.~M.,  {Armstrong} R.,  {Krawiec} C.,   {March} M.~C.,  2016,
  \mn@doi [\mnras] {10.1093/mnras/stw879}, \href
  {http://adsabs.harvard.edu/abs/2016MNRAS.459.4467B} {459, 4467}

\bibitem[\protect\citeauthoryear{{Blazek}, {Mandelbaum}, {Seljak}  \&
  {Nakajima}}{{Blazek} et~al.}{2012}]{Blazek2012}
{Blazek} J.,  {Mandelbaum} R.,  {Seljak} U.,   {Nakajima} R.,  2012, \mn@doi
  [\jcap] {10.1088/1475-7516/2012/05/041}, \href
  {http://adsabs.harvard.edu/abs/2012JCAP...05..041B} {5, 041}

\bibitem[\protect\citeauthoryear{{Blazek}, {Vlah}  \& {Seljak}}{{Blazek}
  et~al.}{2015}]{Blazek2015}
{Blazek} J.,  {Vlah} Z.,   {Seljak} U.,  2015, \mn@doi [\jcap]
  {10.1088/1475-7516/2015/08/015}, \href
  {http://adsabs.harvard.edu/abs/2015JCAP...08..015B} {8, 015}

\bibitem[\protect\citeauthoryear{{Blazek}, {MacCrann}, {Troxel}  \&
  {Fang}}{{Blazek} et~al.}{2017}]{Blazek2017}
{Blazek} J.,  {MacCrann} N.,  {Troxel} M.~A.,   {Fang} X.,  2017, preprint,
  \href {http://adsabs.harvard.edu/abs/2017arXiv170809247B} {} (\mn@eprint
  {arXiv} {1708.09247})

\bibitem[\protect\citeauthoryear{{Bridle} \& {King}}{{Bridle} \&
  {King}}{2007}]{Bridle2007}
{Bridle} S.,  {King} L.,  2007, \mn@doi [New Journal of Physics]
  {10.1088/1367-2630/9/12/444}, \href
  {http://adsabs.harvard.edu/abs/2007NJPh....9..444B} {9, 444}

\bibitem[\protect\citeauthoryear{{Catelan}, {Kamionkowski}  \&
  {Blandford}}{{Catelan} et~al.}{2001}]{Catelan2001}
{Catelan} P.,  {Kamionkowski} M.,   {Blandford} R.~D.,  2001, \mn@doi [\mnras]
  {10.1046/j.1365-8711.2001.04105.x}, \href
  {http://adsabs.harvard.edu/abs/2001MNRAS.320L...7C} {320, L7}

\bibitem[\protect\citeauthoryear{{Chang} et~al.,}{{Chang}
  et~al.}{2013}]{Chang2013}
{Chang} C.,  et~al., 2013, \mn@doi [\mnras] {10.1093/mnras/stt1156}, \href
  {http://adsabs.harvard.edu/abs/2013MNRAS.434.2121C} {434, 2121}

\bibitem[\protect\citeauthoryear{{Chisari}, {Mandelbaum}, {Strauss}, {Huff}  \&
  {Bahcall}}{{Chisari} et~al.}{2014}]{Chisari2014}
{Chisari} N.~E.,  {Mandelbaum} R.,  {Strauss} M.~A.,  {Huff} E.~M.,   {Bahcall}
  N.~A.,  2014, \mn@doi [\mnras] {10.1093/mnras/stu1786}, \href
  {http://adsabs.harvard.edu/abs/2014MNRAS.445..726C} {445, 726}

\bibitem[\protect\citeauthoryear{{Chisari} et~al.,}{{Chisari}
  et~al.}{2015}]{Chisari2015}
{Chisari} N.,  et~al., 2015, \mn@doi [\mnras] {10.1093/mnras/stv2154}, \href
  {http://adsabs.harvard.edu/abs/2015MNRAS.454.2736C} {454, 2736}

\bibitem[\protect\citeauthoryear{{Chisari}, {Dvorkin}, {Schmidt}  \&
  {Spergel}}{{Chisari} et~al.}{2016}]{Chisari2016}
{Chisari} N.~E.,  {Dvorkin} C.,  {Schmidt} F.,   {Spergel} D.~N.,  2016,
  \mn@doi [\prd] {10.1103/PhysRevD.94.123507}, \href
  {http://adsabs.harvard.edu/abs/2016PhRvD..94l3507C} {94, 123507}

\bibitem[\protect\citeauthoryear{{DES Collaboration} et~al.,}{{DES
  Collaboration} et~al.}{2017}]{DES2017}
{DES Collaboration} et~al., 2017, preprint, \href
  {http://adsabs.harvard.edu/abs/2017arXiv170801530D} {} (\mn@eprint {arXiv}
  {1708.01530})

\bibitem[\protect\citeauthoryear{{DESI Collaboration} et~al.,}{{DESI
  Collaboration} et~al.}{2016}]{DESIExperiment}
{DESI Collaboration} et~al., 2016, preprint, \href
  {http://adsabs.harvard.edu/abs/2016arXiv161100036D} {} (\mn@eprint {arXiv}
  {1611.00036})

\bibitem[\protect\citeauthoryear{{Faber} et~al.,}{{Faber}
  et~al.}{2007}]{Faber2007}
{Faber} S.~M.,  et~al., 2007, \mn@doi [\apj] {10.1086/519294}, \href
  {http://adsabs.harvard.edu/abs/2007ApJ...665..265F} {665, 265}

\bibitem[\protect\citeauthoryear{Hamilton}{Hamilton}{2000}]{Hamilton2000}
Hamilton A.,  2000, Monthly Notices of the Royal Astronomical Society, 312, 257

\bibitem[\protect\citeauthoryear{{Hearin} et~al.,}{{Hearin}
  et~al.}{2017}]{Hearin2017}
{Hearin} A.~P.,  et~al., 2017, \mn@doi [\aj] {10.3847/1538-3881/aa859f}, \href
  {http://adsabs.harvard.edu/abs/2017AJ....154..190H} {154, 190}

\bibitem[\protect\citeauthoryear{{Hilbert}, {Xu}, {Schneider}, {Springel},
  {Vogelsberger}  \& {Hernquist}}{{Hilbert} et~al.}{2017}]{Hilbert2017}
{Hilbert} S.,  {Xu} D.,  {Schneider} P.,  {Springel} V.,  {Vogelsberger} M.,
  {Hernquist} L.,  2017, \mn@doi [\mnras] {10.1093/mnras/stx482}, \href
  {http://adsabs.harvard.edu/abs/2017MNRAS.468..790H} {468, 790}

\bibitem[\protect\citeauthoryear{{Hirata} \& {Seljak}}{{Hirata} \&
  {Seljak}}{2004}]{Hirata2004}
{Hirata} C.~M.,  {Seljak} U.,  2004, \mn@doi [\prd]
  {10.1103/PhysRevD.70.063526}, \href
  {http://adsabs.harvard.edu/abs/2004PhRvD..70f3526H} {70, 063526}

\bibitem[\protect\citeauthoryear{{Hirata} et~al.,}{{Hirata}
  et~al.}{2004}]{Hirata2004b}
{Hirata} C.~M.,  et~al., 2004, \mn@doi [\mnras]
  {10.1111/j.1365-2966.2004.08090.x}, \href
  {http://adsabs.harvard.edu/abs/2004MNRAS.353..529H} {353, 529}

\bibitem[\protect\citeauthoryear{{Huff} \& {Mandelbaum}}{{Huff} \&
  {Mandelbaum}}{2017}]{Huff2017}
{Huff} E.,  {Mandelbaum} R.,  2017, preprint, \href
  {http://adsabs.harvard.edu/abs/2017arXiv170202600H} {} (\mn@eprint {arXiv}
  {1702.02600})

\bibitem[\protect\citeauthoryear{{Jeong}, {Komatsu}  \& {Jain}}{{Jeong}
  et~al.}{2009}]{Jeong2009}
{Jeong} D.,  {Komatsu} E.,   {Jain} B.,  2009, \mn@doi [\prd]
  {10.1103/PhysRevD.80.123527}, \href
  {http://adsabs.harvard.edu/abs/2009PhRvD..80l3527J} {80, 123527}

\bibitem[\protect\citeauthoryear{{Joachimi} \& {Schneider}}{{Joachimi} \&
  {Schneider}}{2008}]{Joachimi2008}
{Joachimi} B.,  {Schneider} P.,  2008, \mn@doi [\aap]
  {10.1051/0004-6361:200809971}, \href
  {http://adsabs.harvard.edu/abs/2008A%26A...488..829J} {488, 829}

\bibitem[\protect\citeauthoryear{{Joachimi} \& {Schneider}}{{Joachimi} \&
  {Schneider}}{2010}]{Joachimi2010}
{Joachimi} B.,  {Schneider} P.,  2010, \mn@doi [\aap]
  {10.1051/0004-6361/201014482}, \href
  {http://adsabs.harvard.edu/abs/2010A%26A...517A...4J} {517, A4}

\bibitem[\protect\citeauthoryear{{Joachimi}, {Mandelbaum}, {Abdalla}  \&
  {Bridle}}{{Joachimi} et~al.}{2011}]{Joachimi2011}
{Joachimi} B.,  {Mandelbaum} R.,  {Abdalla} F.~B.,   {Bridle} S.~L.,  2011,
  \mn@doi [\aap] {10.1051/0004-6361/201015621}, \href
  {http://adsabs.harvard.edu/abs/2011A%26A...527A..26J} {527, A26}

\bibitem[\protect\citeauthoryear{{Joachimi} et~al.,}{{Joachimi}
  et~al.}{2015}]{Joachimi2015}
{Joachimi} B.,  et~al., 2015, \mn@doi [\ssr] {10.1007/s11214-015-0177-4}, \href
  {http://adsabs.harvard.edu/abs/2015SSRv..193....1J} {193, 1}

\bibitem[\protect\citeauthoryear{Jones, Oliphant, Peterson  et~al.}{Jones
  et~al.}{2001}]{scipy}
Jones E.,  Oliphant T.,  Peterson P.,   et~al., 2001, {SciPy}: Open source
  scientific tools for {Python}, \url {http://www.scipy.org/}

\bibitem[\protect\citeauthoryear{{Kazin} et~al.,}{{Kazin}
  et~al.}{2010}]{Kazin2010}
{Kazin} E.~A.,  et~al., 2010, \mn@doi [\apj] {10.1088/0004-637X/710/2/1444},
  \href {http://adsabs.harvard.edu/abs/2010ApJ...710.1444K} {710, 1444}

\bibitem[\protect\citeauthoryear{{Kiessling} et~al.,}{{Kiessling}
  et~al.}{2015}]{Kiessling2015}
{Kiessling} A.,  et~al., 2015, \mn@doi [\ssr] {10.1007/s11214-015-0203-6},
  \href {http://adsabs.harvard.edu/abs/2015SSRv..193...67K} {193, 67}

\bibitem[\protect\citeauthoryear{{Kirk} et~al.,}{{Kirk}
  et~al.}{2015}]{Kirk2015}
{Kirk} D.,  et~al., 2015, \mn@doi [\ssr] {10.1007/s11214-015-0213-4}, \href
  {http://adsabs.harvard.edu/abs/2015SSRv..193..139K} {193, 139}

\bibitem[\protect\citeauthoryear{{Krause}, {Eifler}  \& {Blazek}}{{Krause}
  et~al.}{2016}]{Krause2015}
{Krause} E.,  {Eifler} T.,   {Blazek} J.,  2016, \mn@doi [\mnras]
  {10.1093/mnras/stv2615}, \href
  {http://adsabs.harvard.edu/abs/2016MNRAS.456..207K} {456, 207}

\bibitem[\protect\citeauthoryear{{LSST DESC}}{{LSST DESC}}{2017}]{CCLrepo}
{LSST DESC} 2017, LSST DESC Core Cosmology Library,
  \url{https://github.com/LSSTDESC/CCL}

\bibitem[\protect\citeauthoryear{{LSST Science Collaboration} et~al.,}{{LSST
  Science Collaboration} et~al.}{2009}]{LSSTScienceBook2}
{LSST Science Collaboration} et~al., 2009, preprint, \href
  {http://adsabs.harvard.edu/abs/2009arXiv0912.0201L} {} (\mn@eprint {arXiv}
  {0912.0201})

\bibitem[\protect\citeauthoryear{{Laureijs} et~al.,}{{Laureijs}
  et~al.}{2011}]{EuclidRedBook}
{Laureijs} R.,  et~al., 2011, preprint, \href
  {http://adsabs.harvard.edu/abs/2011arXiv1110.3193L} {} (\mn@eprint {arXiv}
  {1110.3193})

\bibitem[\protect\citeauthoryear{{Lesgourgues}}{{Lesgourgues}}{2011}]{Lesgourges2011}
{Lesgourgues} J.,  2011, preprint, \href
  {http://adsabs.harvard.edu/abs/2011arXiv1104.2932L} {} (\mn@eprint {arXiv}
  {1104.2932})

\bibitem[\protect\citeauthoryear{{Lima}, {Cunha}, {Oyaizu}, {Frieman}, {Lin}
  \& {Sheldon}}{{Lima} et~al.}{2008}]{Lima2008}
{Lima} M.,  {Cunha} C.~E.,  {Oyaizu} H.,  {Frieman} J.,  {Lin} H.,   {Sheldon}
  E.~S.,  2008, \mn@doi [\mnras] {10.1111/j.1365-2966.2008.13510.x}, \href
  {http://adsabs.harvard.edu/abs/2008MNRAS.390..118L} {390, 118}

\bibitem[\protect\citeauthoryear{{Loveday} et~al.,}{{Loveday}
  et~al.}{2012}]{Loveday2012}
{Loveday} J.,  et~al., 2012, \mn@doi [\mnras]
  {10.1111/j.1365-2966.2011.20111.x}, \href
  {http://adsabs.harvard.edu/abs/2012MNRAS.420.1239L} {420, 1239}

\bibitem[\protect\citeauthoryear{{Mandelbaum}}{{Mandelbaum}}{2017}]{Mandelbaum2017}
{Mandelbaum} R.,  2017, preprint, \href
  {http://adsabs.harvard.edu/abs/2017arXiv171003235M} {} (\mn@eprint {arXiv}
  {1710.03235})

\bibitem[\protect\citeauthoryear{{Mandelbaum}, {Seljak}  \&
  {Hirata}}{{Mandelbaum} et~al.}{2008}]{Mandelbaum2008}
{Mandelbaum} R.,  {Seljak} U.,   {Hirata} C.~M.,  2008, \mn@doi [\jcap]
  {10.1088/1475-7516/2008/08/006}, \href
  {http://adsabs.harvard.edu/abs/2008JCAP...08..006M} {8, 006}

\bibitem[\protect\citeauthoryear{{Mandelbaum} et~al.,}{{Mandelbaum}
  et~al.}{2011}]{Mandelbaum2011}
{Mandelbaum} R.,  et~al., 2011, \mn@doi [\mnras]
  {10.1111/j.1365-2966.2010.17485.x}, \href
  {http://adsabs.harvard.edu/abs/2011MNRAS.410..844M} {410, 844}

\bibitem[\protect\citeauthoryear{{Mandelbaum}, {Slosar}, {Baldauf}, {Seljak},
  {Hirata}, {Nakajima}, {Reyes}  \& {Smith}}{{Mandelbaum}
  et~al.}{2013}]{Mandelbaum2013}
{Mandelbaum} R.,  {Slosar} A.,  {Baldauf} T.,  {Seljak} U.,  {Hirata} C.~M.,
  {Nakajima} R.,  {Reyes} R.,   {Smith} R.~E.,  2013, \mn@doi [\mnras]
  {10.1093/mnras/stt572}, \href
  {http://adsabs.harvard.edu/abs/2013MNRAS.432.1544M} {432, 1544}

\bibitem[\protect\citeauthoryear{{More}, {Miyatake}, {Mandelbaum}, {Takada},
  {Spergel}, {Brownstein}  \& {Schneider}}{{More} et~al.}{2015}]{More2015}
{More} S.,  {Miyatake} H.,  {Mandelbaum} R.,  {Takada} M.,  {Spergel} D.~N.,
  {Brownstein} J.~R.,   {Schneider} D.~P.,  2015, \mn@doi [\apj]
  {10.1088/0004-637X/806/1/2}, \href
  {http://adsabs.harvard.edu/abs/2015ApJ...806....2M} {806, 2}

\bibitem[\protect\citeauthoryear{{Nakajima}, {Mandelbaum}, {Seljak}, {Cohn},
  {Reyes}  \& {Cool}}{{Nakajima} et~al.}{2012}]{Nakajima2011}
{Nakajima} R.,  {Mandelbaum} R.,  {Seljak} U.,  {Cohn} J.~D.,  {Reyes} R.,
  {Cool} R.,  2012, \mn@doi [\mnras] {10.1111/j.1365-2966.2011.20249.x}, \href
  {http://adsabs.harvard.edu/abs/2012MNRAS.420.3240N} {420, 3240}

\bibitem[\protect\citeauthoryear{{Navarro}, {Frenk}  \& {White}}{{Navarro}
  et~al.}{1997}]{Navarro1997}
{Navarro} J.~F.,  {Frenk} C.~S.,   {White} S.~D.~M.,  1997, \mn@doi [\apj]
  {10.1086/304888}, \href {http://adsabs.harvard.edu/abs/1997ApJ...490..493N}
  {490, 493}

\bibitem[\protect\citeauthoryear{{Newman} et~al.,}{{Newman}
  et~al.}{2015}]{Newman2015}
{Newman} J.~A.,  et~al., 2015, \mn@doi [Astroparticle Physics]
  {10.1016/j.astropartphys.2014.06.007}, \href
  {http://adsabs.harvard.edu/abs/2015APh....63...81N} {63, 81}

\bibitem[\protect\citeauthoryear{{Reid} \& {Spergel}}{{Reid} \&
  {Spergel}}{2009}]{Reid2009}
{Reid} B.~A.,  {Spergel} D.~N.,  2009, \mn@doi [\apj]
  {10.1088/0004-637X/698/1/143}, \href
  {http://adsabs.harvard.edu/abs/2009ApJ...698..143R} {698, 143}

\bibitem[\protect\citeauthoryear{{Reyes}, {Mandelbaum}, {Gunn}, {Nakajima},
  {Seljak}  \& {Hirata}}{{Reyes} et~al.}{2012}]{Reyes2012}
{Reyes} R.,  {Mandelbaum} R.,  {Gunn} J.~E.,  {Nakajima} R.,  {Seljak} U.,
  {Hirata} C.~M.,  2012, \mn@doi [\mnras] {10.1111/j.1365-2966.2012.21472.x},
  \href {http://adsabs.harvard.edu/abs/2012MNRAS.425.2610R} {425, 2610}

\bibitem[\protect\citeauthoryear{{Schechter}}{{Schechter}}{1976}]{Schechter1976}
{Schechter} P.,  1976, \mn@doi [\apj] {10.1086/154079}, \href
  {http://adsabs.harvard.edu/abs/1976ApJ...203..297S} {203, 297}

\bibitem[\protect\citeauthoryear{{Schmidt}, {Newman}, {Abate}  \& {the
  Spectroscopic Needs White Paper Team}}{{Schmidt} et~al.}{2014}]{Schmidt2014}
{Schmidt} S.~J.,  {Newman} J.~A.,  {Abate} A.,   {the Spectroscopic Needs White
  Paper Team} 2014, preprint, \href
  {http://adsabs.harvard.edu/abs/2014arXiv1410.4506S} {} (\mn@eprint {arXiv}
  {1410.4506})

\bibitem[\protect\citeauthoryear{{Schneider} \& {Bridle}}{{Schneider} \&
  {Bridle}}{2010}]{Schneider2010}
{Schneider} M.~D.,  {Bridle} S.,  2010, \mn@doi [\mnras]
  {10.1111/j.1365-2966.2009.15956.x}, \href
  {http://adsabs.harvard.edu/abs/2010MNRAS.402.2127S} {402, 2127}

\bibitem[\protect\citeauthoryear{{Schneider} et~al.,}{{Schneider}
  et~al.}{2013}]{Schneider2013}
{Schneider} M.~D.,  et~al., 2013, \mn@doi [\mnras] {10.1093/mnras/stt855},
  \href {http://adsabs.harvard.edu/abs/2013MNRAS.433.2727S} {433, 2727}

\bibitem[\protect\citeauthoryear{{Sheldon} \& {Huff}}{{Sheldon} \&
  {Huff}}{2017}]{Sheldon2017}
{Sheldon} E.~S.,  {Huff} E.~M.,  2017, \mn@doi [\apj]
  {10.3847/1538-4357/aa704b}, \href
  {http://adsabs.harvard.edu/abs/2017ApJ...841...24S} {841, 24}

\bibitem[\protect\citeauthoryear{{Sheldon} et~al.,}{{Sheldon}
  et~al.}{2004}]{Sheldon2004}
{Sheldon} E.~S.,  et~al., 2004, \mn@doi [\aj] {10.1086/383293}, \href
  {http://adsabs.harvard.edu/abs/2004AJ....127.2544S} {127, 2544}

\bibitem[\protect\citeauthoryear{{Sif{\'o}n}, {Hoekstra}, {Cacciato}, {Viola},
  {K{\"o}hlinger}, {van der Burg}, {Sand}  \& {Graham}}{{Sif{\'o}n}
  et~al.}{2015}]{Sifon2015}
{Sif{\'o}n} C.,  {Hoekstra} H.,  {Cacciato} M.,  {Viola} M.,  {K{\"o}hlinger}
  F.,  {van der Burg} R.~F.~J.,  {Sand} D.~J.,   {Graham} M.~L.,  2015, \mn@doi
  [\aap] {10.1051/0004-6361/201424435}, \href
  {http://adsabs.harvard.edu/abs/2015A%26A...575A..48S} {575, A48}

\bibitem[\protect\citeauthoryear{{Singh} \& {Mandelbaum}}{{Singh} \&
  {Mandelbaum}}{2016}]{Singh2016}
{Singh} S.,  {Mandelbaum} R.,  2016, \mn@doi [\mnras] {10.1093/mnras/stw144},
  \href {http://adsabs.harvard.edu/abs/2016MNRAS.457.2301S} {457, 2301}

\bibitem[\protect\citeauthoryear{{Singh}, {Mandelbaum}  \& {More}}{{Singh}
  et~al.}{2015}]{Singh2014}
{Singh} S.,  {Mandelbaum} R.,   {More} S.,  2015, \mn@doi [\mnras]
  {10.1093/mnras/stv778}, \href
  {http://adsabs.harvard.edu/abs/2015MNRAS.450.2195S} {450, 2195}

\bibitem[\protect\citeauthoryear{{Singh}, {Mandelbaum}, {Seljak}, {Slosar}  \&
  {Vazquez Gonzalez}}{{Singh} et~al.}{2016}]{Singh2016b}
{Singh} S.,  {Mandelbaum} R.,  {Seljak} U.,  {Slosar} A.,   {Vazquez Gonzalez}
  J.,  2016, preprint, \href
  {http://adsabs.harvard.edu/abs/2016arXiv161100752S} {} (\mn@eprint {arXiv}
  {1611.00752})

\bibitem[\protect\citeauthoryear{{Smith} et~al.,}{{Smith}
  et~al.}{2003}]{Smith2003}
{Smith} R.~E.,  et~al., 2003, \mn@doi [\mnras]
  {10.1046/j.1365-8711.2003.06503.x}, \href
  {http://adsabs.harvard.edu/abs/2003MNRAS.341.1311S} {341, 1311}

\bibitem[\protect\citeauthoryear{{Spergel} et~al.,}{{Spergel}
  et~al.}{2015}]{WFIRSTReport}
{Spergel} D.,  et~al., 2015, preprint, \href
  {http://adsabs.harvard.edu/abs/2015arXiv150303757S} {} (\mn@eprint {arXiv}
  {1503.03757})

\bibitem[\protect\citeauthoryear{{Takahashi}, {Sato}, {Nishimichi}, {Taruya}
  \& {Oguri}}{{Takahashi} et~al.}{2012}]{Takahashi2012}
{Takahashi} R.,  {Sato} M.,  {Nishimichi} T.,  {Taruya} A.,   {Oguri} M.,
  2012, \mn@doi [\apj] {10.1088/0004-637X/761/2/152}, \href
  {http://adsabs.harvard.edu/abs/2012ApJ...761..152T} {761, 152}

\bibitem[\protect\citeauthoryear{{Tenneti}, {Mandelbaum}, {Di Matteo}, {Feng}
  \& {Khandai}}{{Tenneti} et~al.}{2014}]{Tenneti2014}
{Tenneti} A.,  {Mandelbaum} R.,  {Di Matteo} T.,  {Feng} Y.,   {Khandai} N.,
  2014, \mn@doi [\mnras] {10.1093/mnras/stu586}, \href
  {http://adsabs.harvard.edu/abs/2014MNRAS.441..470T} {441, 470}

\bibitem[\protect\citeauthoryear{{Tenneti}, {Mandelbaum}, {Di Matteo},
  {Kiessling}  \& {Khandai}}{{Tenneti} et~al.}{2015}]{Tenneti2015}
{Tenneti} A.,  {Mandelbaum} R.,  {Di Matteo} T.,  {Kiessling} A.,   {Khandai}
  N.,  2015, \mn@doi [\mnras] {10.1093/mnras/stv1625}, \href
  {http://adsabs.harvard.edu/abs/2015MNRAS.453..469T} {453, 469}

\bibitem[\protect\citeauthoryear{{Troxel} \& {Ishak}}{{Troxel} \&
  {Ishak}}{2015}]{Troxel2015}
{Troxel} M.~A.,  {Ishak} M.,  2015, \mn@doi [\physrep]
  {10.1016/j.physrep.2014.11.001}, \href
  {http://adsabs.harvard.edu/abs/2015PhR...558....1T} {558, 1}

\bibitem[\protect\citeauthoryear{{Velliscig} et~al.,}{{Velliscig}
  et~al.}{2015a}]{Velliscig2015}
{Velliscig} M.,  et~al., 2015a, \mn@doi [\mnras] {10.1093/mnras/stv1690}, \href
  {http://adsabs.harvard.edu/abs/2015MNRAS.453..721V} {453, 721}

\bibitem[\protect\citeauthoryear{{Velliscig} et~al.,}{{Velliscig}
  et~al.}{2015b}]{Velliscig2015b}
{Velliscig} M.,  et~al., 2015b, \mn@doi [\mnras] {10.1093/mnras/stv2198}, \href
  {http://adsabs.harvard.edu/abs/2015MNRAS.454.3328V} {454, 3328}

\bibitem[\protect\citeauthoryear{{Weinberg}, {Mortonson}, {Eisenstein},
  {Hirata}, {Riess}  \& {Rozo}}{{Weinberg} et~al.}{2013}]{Weinberg2013}
{Weinberg} D.~H.,  {Mortonson} M.~J.,  {Eisenstein} D.~J.,  {Hirata} C.,
  {Riess} A.~G.,   {Rozo} E.,  2013, \mn@doi [\physrep]
  {10.1016/j.physrep.2013.05.001}, \href
  {http://adsabs.harvard.edu/abs/2013PhR...530...87W} {530, 87}

\bibitem[\protect\citeauthoryear{{Zu} \& {Mandelbaum}}{{Zu} \&
  {Mandelbaum}}{2015}]{Zu2015}
{Zu} Y.,  {Mandelbaum} R.,  2015, \mn@doi [\mnras] {10.1093/mnras/stv2062},
  \href {http://adsabs.harvard.edu/abs/2015MNRAS.454.1161Z} {454, 1161}

\bibitem[\protect\citeauthoryear{{van Uitert} et~al.,}{{van Uitert}
  et~al.}{2017}]{vanUitert2017}
{van Uitert} E.,  et~al., 2017, preprint, \href
  {http://adsabs.harvard.edu/abs/2017arXiv170605004V} {} (\mn@eprint {arXiv}
  {1706.05004})

\bibitem[\protect\citeauthoryear{van~der Walt, Colbert  \& Varoquaux}{van~der
  Walt et~al.}{2011}]{numpy}
van~der Walt S.,  Colbert S.~C.,   Varoquaux G.,  2011, Computing in Science
  and Engineering, 13

\makeatother
\end{thebibliography}
\end{document}